\documentclass[%
 reprint,
 amsmath,amssymb,
 aps,
floatfix,
]{revtex4-2}

\usepackage{graphicx}
\usepackage{dcolumn}
\usepackage{bm}
\usepackage{xcolor}
\usepackage{braket}
\usepackage{mathtools}
\usepackage{makecell}
\usepackage{multirow}
\usepackage{array}
\usepackage{booktabs}
\usepackage{collcell}
\usepackage{amsthm}

\usepackage{float}
\usepackage{tabularx}
\usepackage{cellspace} 
\newcommand{\ketbra}[2]{\mathinner{|{#1}\rangle\langle{#2}|}}
\newcolumntype{Y}{>{\centering\arraybackslash}X}
\newcolumntype{M}{>{$\displaystyle}Y<{$}} 

\newcommand{\eqnumber}[1]{\refstepcounter{equation}\label{#1}(\theequation)}

\usepackage{graphicx}
\usepackage{hyperref}

\newtheorem{misconception}{Misconception}

\begin{document}

\preprint{APS/123-QED}

\title{Beyond the Carnot limit: work extraction via an entropy battery}

\author{Liam J McClelland}
\email{liam.mcclelland@griffithuni.edu.au}

\affiliation{Queensland Quantum and Advanced Technologies Research Institute, Griffith University, Brisbane, Queensland 4111 Australia }

\date{\today}

\begin{abstract}
Heat is a physical manifestation of entropy, where the removal of entropy from a thermal energy reservoir permits the conversion of heat into work. This entropy transfer is facilitated by the cold thermal energy reservoir in typical heat engines. Recent developments in quantum heat engines that operate between thermal energy and spin angular momentum reservoirs show that it is possible to transfer entropy out of energy and into a different conserved quantity. The implications of this type of entropy transfer have not been fully explored, especially on the work extractable using an ensemble with multiple conserved quantities. Using the aforementioned heat engines, we show that such an ensemble will transform heat into work beyond the Carnot efficiency limit while operating at maximum power. This result is obtained without induced quantum coherence, a technique commonly used in the field of quantum heat engines to achieve the same outcome. Without loss of generality, we also show that thermal spin reservoirs behave as thermodynamic baths with well-defined temperatures, heat capacities, and fluctuation–dissipation relations. Finally, our analysis of entropy capacity suggests that particle indistinguishability is necessary for inter-particle interactions, and for entropy to transfer between canonical ensembles. These results establish a foundation for entropy-based quantum devices that can extract work from a hot thermal energy reservoir more efficiently than possible with a cold thermal energy reservoir. These devices also act as high energy-density batteries and efficient heat storage systems. Our results will have implications for quantum heat engines, spinor condensates, spintronics, and quantum batteries.
\end{abstract}

\maketitle


\section{Introduction}
Heat engines provide not only practical utility, but also a theoretical framework for describing some of the most fundamental principles of nature. The pursuit of increasingly powerful and efficient engines, refrigerators, and related devices, driven by both technological ambition and the intellectual curiosity of figures such as Carnot and Clausius \cite{carnot1872reflexions}, led to the formulation of the four laws of thermodynamics. The need to explain the emergence of these laws from the microscopic constituents of thermodynamic systems lead to statistical mechanics, a theory that successfully described the macroscopic properties of the ensemble using probability and statistical techniques. Maxwell, by utilizing statistical mechanics, proposed a mechanism by which an entity with sufficient knowledge of the microscopic states could violate the second law \cite{maxwell18711902}, in what came to be known as Maxwell's demon. 

Szilard \cite{szilard1964uber}, Landauer \cite{Landauer}, and Bennett \cite{bennett2003notes} each contributed, at times indirectly, to the resolution of the demon paradox, with Bennett ultimately recognising that such a demon would only operate for as long as it's memory has capacity to store information about the state of the system. If this memory capacity were saturated, as would occur for a demon constrained by finite physical resources, no further work could be extracted from the system. Bennett showed, using Landauer's principal, that the energetic cost of erasing the demon's memory would be greater-than or equal-to the amount of work extracted during the engine cycles, therefore closing the paradox.

The roles of Information and entropy (the latter being the lack of the former) in thermodynamics became a recurring theme in the study of thermodynamic systems, even prior to the resolution of the demon paradox. Jaynes, from the perspective of information theory, generalized statistical mechanics to include arbitrary conserved quantities, therefore removing the special treatment given to energy \cite{jaynes1957information, jaynes1957information2}. His rational can be intuitively understood by recognizing that the time invariant nature of conserved quantities must allow for the encoding of information. Vaccaro and Barnett later used Jaynes's principle to show how the erasure of information in the Maxwell demon thought experiment can be performed using spin angular momentum \cite{Vaccaro_2011, Croucher_2018, croucher2017discrete, croucher2021thermodynamics}, leading to the development of quantum spin heat engines \cite{writeQHEoperating2018, andre2025QSHE} and gas spin heat engines \cite{manakil2025GSHE}. The application of Jaynes's work to quantum systems has also led to developments in generalized heat baths, quantum batteries, and resource theories \cite{Guryanova_2016,Yunger_Halpern_2016, chitambar2019quantum}. It would now appear as though thermodynamics is not describing energy and heat, but instead information and entropy. However, unlike energy and heat, information and entropy can transfer between conserved quantities according to \cite{Vaccaro_2011, Croucher_2018, croucher2017discrete, croucher2021thermodynamics, writeQHEoperating2018, andre2025QSHE, manakil2025GSHE, Guryanova_2016, Yunger_Halpern_2016}, where the conserved quantities act as the physical degrees of freedom for which information is encoded. The concept of freely transferrable entropy across conserved quantities has not been widely utilized by the scientific community as no clear practicality has emerged for applying thermodynamics in this manner. This article provides one such practicality.

This article is structured in the following manner. In \S \ref{sec: Entropy as a conserved quantity} we discuss our rational for treating entropy as a conserved quantity in isolated systems. In \S \ref{sec: The application of quantum information} we discuss the current applications of quantum information and the open question we are investigating. In \S\ref{sec: The entropy battery} we introduce the entropy battery, followed by \S \ref{sec: Results} we present the energy efficiency of the battery. In this section we also summarize the thermodynamic properties of spin angular momentum, which are derived in the Appendix. In \S \ref{sec: Discussion} we discuss the implications of this type of device, and finally, in \S \ref{sec: Maximising entropy capcity} we present methods of maximizing the batteries entropy capacity.

\section{Entropy as a conserved quantity}\label{sec: Entropy as a conserved quantity}
In quantum mechanics, isolated systems evolve unitarily and the system's total entropy is conserved $\delta\mathbb{S}=0$. Subsystem interactions do not produce entropy, even when those subsystems are at non-equilibrium. This fact is formalized in the `no-hiding' \cite{PhysRevLett.98.080502}, `no-cloning' \cite{wootters1982single}, and `no-deletion' \cite{kumar2000impossibility} theorems, all of which point to the impossibility of creating or destroying quantum information, with early experimental evidence \cite{PhysRevLett.106.080401, Gao_2022} agreeing. Despite the evidence of the conservation of information and entropy, there is often misconceptions of whether entropy production can occur in isolated systems. This usually stems from the original formulation of the second law $\delta\mathbb{S}\ge 0$, and reinforced by Boltzmann's statistics and associated entropy \cite{lebowitz1993boltzmann}. However, as already discussed, no such entropy production can occur if the system is isolated. Any perception of entropy production in quantum systems is due to two common misconceptions,
\begin{misconception}
  A system's total entropy after subsystem interaction can be calculated as a sum of the subsystem entropies.
\end{misconception}
\begin{misconception}
  An isolated system can decohere, and therefore a systems entropy can increase.
\end{misconception}
Misconception 1 is usually made if subsystems are assumed to be independent at all times. If the subsystems interact, with interactions necessarily correlating (entangling) the subsystems, then this assumption is clearly false. Entanglement leads to the inseparability of subsystem wavefunctions and entropy sub-additivity  $\sum_i\delta\mathbb{S}_i\ge\delta\mathbb{S}=0$, where $\mathbb{S}_i=-\text{Tr}[\rho_i\ln(\rho_i)]$ is the entropy of subsystem $i$ with reduced density matrix $\rho_i=\text{Tr}_j[\rho]$, and $\rho$ is the system density matrix. The sub-additivity of subsystem entropies creates the appearance of entropy production in the total system. Without misconception 1, we return to the true change in system entropy $\delta\mathbb{S}=0$. 

Misconception 2, although less common than 1, comes from a misunderstanding as to the origin of decoherence. Decoherence only occurs when a system interacts with an external environment, and it has been shown both theoretically \cite{RevModPhys.75.715} and to some extent experimentally \cite{PhysRevLett.106.080401} that unitary evolution of system+environment creates the appearance of decoherence in the system. With decoherence typically leaving the system in a thermal state after a sufficiently long time, one often incorrectly concludes that entropy production can occur in isolated, decohering systems.

There are also cases in classical mechanics where entropy production causes inconsistencies. The Gibbs paradox, although seemingly unrelated, points to the issue of entropy production for two ideal gases that are arbitrarily similar. This problem is only resolved once particle indistinguishability is accepted \cite{SAUNDERS202037}. Particle indistinguishability relates directly to the permutation of information (a parity operation), or equally to the transferability of information between particles. The resolution to the Gibbs paradox, along with the requirement of photon indistinguishability to resolve the UV catastrophe, the indistinguishability of Bosons and Fermions in statistical mechanics, the behavior of heat capacities at low temperature, and inherent indistinguishability in quantum mechanics suggests that entropy production in macroscopic systems is also illusionary. Instead, for all of the evidence to be consistent, then entropy and information must be conserved, with them only able to transfer between the subsystem of an isolated system. Whether the true nature entropy of as an `anthropomorphic' by-product of our ignorance as an observer \cite{jaynes1965gibbs}, or is as physical as the conserved quantities themselves, the act of obtaining information and reducing the systems entropy takes physical resources \cite{bennett1998quantum, Croucher_2018, Vaccaro_2011}, then entropy must be treated as a resource itself.

For these reasons, we treat information and entropy as conserved quantities, with entropy and information manifesting in thermodynamic systems as generalized heats and works. However, while physical conserved quantities such as energy are constrained to their domains, entropy and information can in principle reversibly traverse the physical conserved quantities \cite{Croucher_2018, Vaccaro_2011, Guryanova_2016, writeQHEoperating2018, andre2025QSHE, manakil2025GSHE}, meaning the generalized heats and works in each of the conserved quantities can reversibly exchange. We will discuss this further in \S \ref{sec: The entropy battery}. This idea points to an intriguing possibility: the exchange of entropy from a conserved quantity we wish to utilize, such as energy, into another conserved quantity that can be discarded, such as spin.

\section{The application of quantum information}\label{sec: The application of quantum information}
The fields of modern resource theories, quantum batteries, and quantum heat engines all make use of quantum information to varying extent. They provide pathways towards quantum advantage in quantum thermodynamic systems. These advantages include: collective power enhancement in the charging of quantum batteries \cite{PhysRevLett.120.117702}, heat engines utilizing non-energetic cold baths \cite{andre2025QSHE, writeQHEoperating2018, manakil2025GSHE}, non-classical and non-energetic methods of information erasure \cite{Croucher_2018, Vaccaro_2011}, the saturation and surpassing of the classical Carnot energy efficiency limit \cite{bender2000quantum,rossnagel2014nanoscale, PhysRevX.7.031044, PhysRevE.86.051105, PhysRevE.93.052120, correa2014quantum, scully2003extracting, dillenschneider2009energetics, PhysRevE.76.031105, PhysRevLett.123.240601, annurev:/content/journals/10.1146/annurev-physchem-040513-103724}, and methods of error correction in qubits \cite{fernandez2004algorithmic, doi:10.1073/pnas.241641898}. Information and resource theories have separately found broad application in the fields of quantum state amplification and metrology \cite{marvian2014extending, Marvian_2016, caves1982quantum}. This list of applications, although hardly exhaustive, does point to some of the main ways quantum information principles are used in real systems. 

It should be noted that the observation of super-Carnot efficiencies in certain quantum heat engines has only been achieved by using modified (squeezed, induced coherence) thermal energy reservoirs, with all such engines operating below the generalised Carnot limit \cite{Abah_2014}. These non-thermal states require resources (typically energy) to prepare. This initial resource cost reduces the total efficiency of the heat engine below the Carnot limit. It is our assessment that, except for heat engines that utilize multiple conserved quantities (spin angular momentum and energy in the case of \cite{andre2025QSHE,writeQHEoperating2018}), no quantum heat engine has truly surpassed the Carnot limit. 

As for the transfer of entropy between conserved quantities, \cite{Croucher_2018, croucher2021thermodynamics,croucher2017discrete, Guryanova_2016} have independently explored the reversible exchange of conserved quantity entropies within the framework of generalized thermodynamics. Meanwhile, \cite{writeQHEoperating2018, andre2025QSHE, manakil2025GSHE} provide explicit unitary operations that couple spin angular momentum and energy, operations that we later use in \S \ref{sec: The entropy battery} for the entropy battery. These studies have primarily focused on the principles of entropy transfer or the actual heat engines that run between two conserved quantities (energy and spin) on single quantum systems (quantum dots or trapped ions). Therefore, to the best of our knowledge, there has been very little exploration of an \textit{ensemble's} capacity to efficiently store entropy over \textit{multiple} conserved quantities, and the effect this has on the efficiency of energetic work extraction.

There are eight conserved quantities for which information can in principle be encoded, one for each physical symmetry according to Noether’s theorem \cite{Noether_1971, Kosmann-Schwarzbach2011}. The known conserved quantities are presented in Table \ref{tab:conserved quantities}.
\begin{table}
    \centering
    \begin{tabularx}{\linewidth}{Sl|Y|Y}
        \hline\hline
        \textbf{Conserved quantity} & \textbf{\makecell{`Heat'\\analogue}} & \textbf{\makecell{`Work'\\analogue}}\\
        \hline
        Energy & heat $Q$ & work $W$\\
        Spin angular momentum & spin therm $\mathcal{Q}$ & spin labor $\mathcal{L}$\\
        \hline
        Linear momentum &\multirow{5}{*}{\makecell[c]{\\\\Generalized\\heat $\mathbb{Q}$}} &\multirow{5}{*}{\makecell[c]{\\\\Generalized\\work $\mathbb{W}$}}\\
        Lorentz 3-vector boost &&\\
        Electric charge &  & \\
        Color charge &  &  \\
        Weak isospin &  &  \\
        \hline\hline
        
    \end{tabularx}
    \caption{The known conserved quantities as derived from continuous physical symmetries according to Noether's theorem, excluding Baryon, Lepton difference. Also shown are the known analogous definitions of `heat' and `work' for these conserved quantities, with $\mathbb{Q}$ and $\mathbb{W}$ in place for the quantities without literature definitions. Energy, linear momentum, and the Lorentz 3-vector boost may be defined on both discrete and continuous state spaces, while spin angular momentum, electric charge, color charge, and weak isospin are intrinsically discrete and bounded. The generalized heat and work are assumed to exist for these conserved quantities based on the commentary in this paper, although the application of these quantities as useful thermodynamic properties is questionable with the current state of technology.}
    \label{tab:conserved quantities}
\end{table}
Of the eight conserved quantities, only energy and spin angular momentum are accessible and controllable by bench top experiments, with systems such as Bose–Einstein condensates and individually trapped ions allowing controllable preparation of both thermal motion and internal spin states. We will therefore focus our efforts in this article on energy in a single motional axis and the z-component of spin angular momentum, with our analysis applying equally to any conserved quantity.


\section{The entropy battery}\label{sec: The entropy battery}
The entropy battery is a device that capitalizes on entropy conservation, although physically it is no more than a canonical ensemble with multiple conserved quantities. We assume the container for which the particles exist in is a quadratic potential, such that each particle acts as a quantum harmonic oscillator with energy spacing $h\nu$. We then suppose each particle has internal, energy degenerate spin states. These spin states correspond to the quantized values of the $z$-component of spin angular momentum. The canonical ensemble as described therefore contains two conserved quantities, which can be made mutually independent by setting the Zeeman splitting to 0, $\delta_B=0$. Our analysis will continue to apply even for $|\delta_B|>0$, as long as $|\delta_B|=|g\mu_B m_jB_z|$ is much less than the harmonic spacing $h\nu$, or $|B_z|\ll |h\nu/g\mu_Bm_j|$.

What may not be obvious is that these now independent conserved quantities act as independent thermodynamic reservoirs in the energy and spin angular momentum degrees of freedom. In Appendices \ref{sec: Spin temperature}-\ref{sec: Waste response appendix} we explicitly show that an ensemble with only spin angular momentum has thermodynamic properties, including: spin analogous heat, work and heat capacity (re-named spin therm, spin labor and waste capacity respectively), unitless temperature $\tau$, and Bose-Einstein and Fermi-Dirac statistics. These properties and more are presented as a summary in Table.\ref{table: Thermodynamics of spin systems}, with them equally applying to any mutually independent conserved quantity. 
\begin{table*}[p]
    \begin{tabularx}{\linewidth}{@{}cSl|Y|Mr@{}}
    \toprule
    \hline
       &\textbf{Thermodynamic property}& \textbf{Energy} &\textbf{Spin angular momentum}&\\
       \hline
       \multirow{8}{*}{\rotatebox[origin=c]{90}{}}&Entropy&$\displaystyle\mathbb{S}_E=\ln(Z_E)+\beta\braket{E}$  & \mathbb{S}_{s}=\ln(Z_{s})+\gamma\braket{\bar{J}_z}&\eqnumber{eq:Entropy}\\
       &\makecell{Partition function\\ ($\Omega=\text{microstates}$)}&$\displaystyle Z_E=\sum_{\Omega}e^{-\beta E(\Omega)}$&Z_s=\sum_{\Omega}e^{-\gamma J(\Omega)}&\eqnumber{eq: Partition function}\\
       &\makecell[l]{Inverse temperature} &$\displaystyle\beta=\frac{1}{k_BT}$&\gamma=\frac{1}{\tau}&\eqnumber{eq:Temperature}\\
       &\makecell[l]{Entropy-energy/\\spin relation}& $\displaystyle\frac{\partial\mathbb{S}_E}{\partial \braket{E}}=\frac{1}{T}$&\frac{\partial\mathbb{S}_{s}}{\partial \braket{\bar{J}_z}}=\frac{1}{\tau} &\eqnumber{eq: Entropy-momentum relation}\\
       &\makecell[l]{Average conserved quantity}&$\displaystyle\braket{E}=-\frac{\partial}{\partial\beta}\ln(Z_E)=T^2\frac{\partial}{\partial T}\ln(Z_E)$&\braket{\bar{J}_z}=-\frac{\partial}{\partial\gamma}\ln(Z_{s})=\tau^2\frac{\partial}{\partial\tau}\ln(Z_s) &\eqnumber{eq:AverageSpinInTermsOfZ}\\
       &\makecell[l]{Heat capacity/\\waste response}&$\displaystyle C_V=\left(\frac{\partial\braket{E}}{\partial T}\right)_V$ & C_s=\frac{d\braket{\bar{J}_z}}{d\tau}&\eqnumber{eq:Wasteresponse}\\
       &\makecell[l]{Entropic response}&$\displaystyle \frac{1}{T}C_V=\left(\frac{\partial\mathbb{S}_E}{\partial T}\right)_V$ & \frac{1}{\tau}C_s=\frac{d\mathbb{S}_s}{d\tau}&\eqnumber{eq:Entropicresponse}\\
       &\makecell[l]{Heat/spin therm}&$\displaystyle Q=\int C_V dT$ &\mathcal{Q}=\int C_sd\tau&\eqnumber{eq:Heat}\\
       &\makecell[l]{Bosonic heat in the\\thermodynamic limit} & $\displaystyle\frac{Q}{N}=\frac{1}{2}\sum_{j}\varepsilon_j(\coth(\varepsilon_j/2k_BT)-1)$ &\frac{\mathcal{Q}}{N}=\frac{1}{2}\sum_{j=1}^{2S}j\left(\coth(j/2\tau)-1\right)&\eqnumber{eq:Bosonic heat}\\
       &\makecell[l]{Work/spin labor \cite{Croucher_2018}}&$\displaystyle W=Q_1+Q_2$&\mathcal{L}=\mathcal{Q}_1+\mathcal{Q}_2&\eqnumber{eq:Work}\\\hline
       \multirow{4}{*}{\rotatebox[origin=c]{90}{\textbf{Four laws}\textcolor{white}{hi}}}&Zero'th law&$\displaystyle\lim_{T\rightarrow 0}\mathbb{S}_E=0$ & \lim_{\tau\rightarrow 0}\mathbb{S}_{s}=0 &\eqnumber{eq:Zero'thlaw}\\
       &First law (closed system)&$\displaystyle\delta\braket{E}=0$ & \delta\braket{\bar{J}_z}=0 & \eqnumber{eq:Firstlaw}\\
       
       &Second law (closed system)&$\displaystyle\delta \mathbb{S}_E=0$ &\delta \mathbb{S}_{s}=0 &\eqnumber{eq:Second law}\\
       
       &Third law&$\displaystyle T_1=T_2$&\tau_1=\tau_2& \eqnumber{eq:Third law}\\
       \hline
       
       \multirow{6}{*}{\rotatebox[origin=c]{90}{\textbf{Ensemble statistics}\textcolor{white}{hihihihihihihihi}}}&\makecell[l]{Bose-Einstein/Fermi-dirac\\($\mp$) distributions \cite{bose1994man}} &$\displaystyle\braket{n_\varepsilon}=\frac{1}{e^{(\varepsilon - \mu)/k_BT}\mp 1}$&\braket{n_j}=\frac{1}{e^{(j - S)/\tau}\mp 1}&\eqnumber{eq:BoseEinsteinFermiDiracdistributions}\\
       &\makecell[l]{Distinguishable heat\\capacity/waste response \cite{MARCONI_2008}}&$\displaystyle \frac{C_V}{N}=\frac{1}{k_BT^2}\times \text{Var}(E(T))$&\frac{C_s}{N}=\frac{1}{\tau^2}\times \text{Var}(\bar{J}_z(\tau))&\eqnumber{eq:Distinguishablespecificheat}\\
       &\makecell[l]{Einstein Solid model\\per degree of freedom \cite{rogers2005einstein}}&$\displaystyle
\frac{C_V}{N}=\frac{1}{4k_BT^2}\,\times\,\frac{\varepsilon}{\sinh^2(\varepsilon/2k_BT)}
$&\frac{C_s}{N}=\frac{1}{4\tau^2}\,\times\,\frac{1}{\sinh^2(1/2\tau)}&\eqnumber{eq:EinsteinSolid}\\
       &\makecell[l]{Bosonic heat capacity/\\waste response } & $\displaystyle\frac{C_V}{N}=\frac{1}{4k_BT^2}\sum_{j}\frac{\varepsilon_j^2}{\sinh^2(\varepsilon_j/2k_BT)}$ & \frac{C_s}{N}=\frac{1}{4\tau^2}\sum_{j=1}^{2S}\frac{j^2}{\sinh^2(j/2\tau)}&\eqnumber{eq:Bosonicspecificheat}\\
       &\makecell[l]{Debye model\\per degree of freedom \cite{rogers2005einstein}} & $\displaystyle\frac{C_V}{N}=\frac{3}{4k_BT^2 \varepsilon_D^2}\int_{0}^{\varepsilon_D}\frac{\varepsilon_j^2}{\sinh^2(\varepsilon_j/2k_BT)}\varepsilon_j^2d\varepsilon$ & \frac{C_s}{N}=\frac{3}{4\tau^2(2S)^2}\int_{0}^{2S}\frac{j^2}{\sinh^2(j/2\tau)}j^2dj&\eqnumber{eq:Debyemodel}\\
       &\makecell[l]{Distinguishable\\Entropy capacity} &$\mathbb{S}_\text{max}=N\ln(d_E)$ & \mathbb{S}_\text{max}=N\ln(d_s) &\eqnumber{eq:Distinguishable entropy capacity}\\
       &\makecell[l]{Bosonic\\Entropy capacity} &$\mathbb{S}_\text{max}=\ln\begin{pmatrix}N+d_E-1\\d_E-1\end{pmatrix}$ & \mathbb{S}_\text{max}=\ln\begin{pmatrix}N+d_s-1\\d_s-1\end{pmatrix} &\eqnumber{eq:Bosonic entropy capacity}\\
       &\makecell[l]{Fermionic\\Entropy capacity} &$\mathbb{S}_\text{max}=\ln\begin{pmatrix}d_E\\N\end{pmatrix}$ & \mathbb{S}_\text{max}=\ln\begin{pmatrix}d_s\\N\end{pmatrix} &\eqnumber{eq:Fermionic entropy capacity}\\
       \hline\hline
    \end{tabularx}
    \caption{Summary of spin-analogous thermodynamic properties. In these expressions, $\braket{\bar{J}_z}=\braket{J_z}/\hbar$ is the unitless average angular momentum, $d_m$ is the number of states for the energy, spin $m=E,s$ degrees of freedom respectively, $N$ is the number of particles, $\varepsilon_j=h\nu_j$ is the energy of the $j$\textsuperscript{th} vibrational mode with frequency $\nu_j$, and $\varepsilon_D=h\nu_D$ is the Debye energy. Where relevant we have also included references. The waste response is a generalization of the heat capacity to arbitrary conserved quantities, and is coined based on the results of the fluctuation-dissipation theorem, whereby $C_V$ and $C_s$ are response functions \eqref{eq:Distinguishablespecificheat}. The derivation for each of these expressions for the spin angular momentum conserved quantity is presented in appendices \ref{sec: Spin temperature}-\ref{sec: Waste response appendix}. The first and second laws of generalized thermodynamics apply to closed systems, which is the scenario used in the entropy battery model. The third law describes the generalized thermalization condition of two canonical ensembles.}\label{table: Thermodynamics of spin systems}
\end{table*}

As these reservoirs are initially independent, the entropy of the ensemble is extensive
\begin{equation}
\mathbb{S}_\text{battery}=\mathbb{S}_\text{energy}+\mathbb{S}_\text{spin}.\label{eq: Extensive entropy of energy and spin}
\end{equation}
The entropy capacity (entropy at infinite unitless temperature $\tau=\infty$) of each reservoir is determined by eqs.\eqref{eq:Distinguishable entropy capacity}-\eqref{eq:Fermionic entropy capacity}, depending on the particles' statistics. The entropy capacity for a single conserved quantity reservoir is dependent only on the number of microstates $\Omega$, with $\Omega$ dependent on the number of states per particle $d$ and the number of particles $N$. 

\begin{figure}[t]
    \centering
    \includegraphics[width=\linewidth]{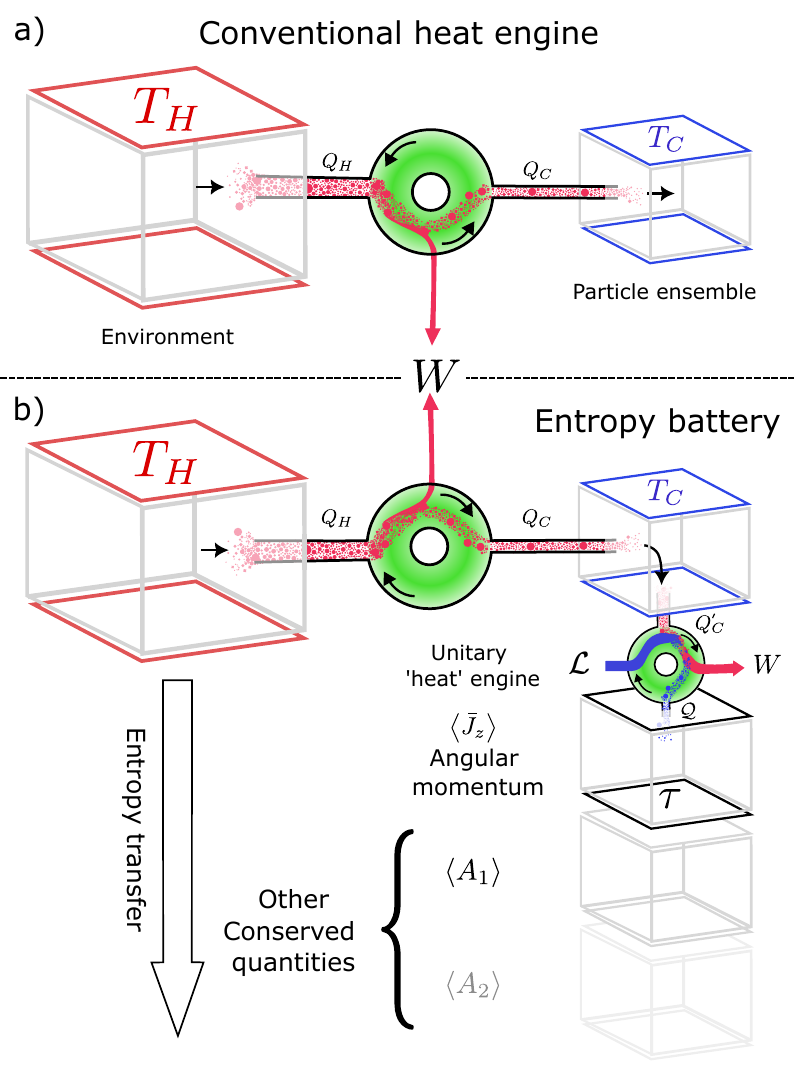}
    \caption{Overview of (a) a conventional Carnot heat engine between hot and cold baths, and (b) the entropy battery. In (b), the equivalent conserved quantity `heat' baths with general conserved quantity $\braket{A_m}$ are represented by black cubes aligned vertically. Energy couples to spin angular momentum within the entropy battery via a unitary `heat engine', realized experimentally by coherent Raman transitions \cite{writeQHEoperating2018, andre2025QSHE, manakil2025GSHE}. Spin labor $\mathcal{L}$ within the entropy battery enables the conversion of heat energy into work. There aren't any equivalent unitary operations for the other conserved quantities that we are aware of. Entropy from the environmental reservoir is transferred into the battery’s internal reservoirs (energy, spin, etc.), enhancing work extraction until their entropy capacities are saturated or equilibrium is reached. As these other conserved quantity `heat' baths reside within the battery’s particles, the entropy battery achieves a higher per particle work extraction density than possible with conventional heat engines.}
    \label{fig: Extensive entropy storage}
\end{figure}

For these two reservoirs to reversibly exchange entropy, we utilize unitary operations such as the coherent Raman transition from \cite{andre2025QSHE,writeQHEoperating2018, manakil2025GSHE}. This allows for an effective thermalizing between the initially independent energy and spin reservoirs. Much like regular adiabatic heat engines, this thermalization produces useful work. Unlike regular heat engines, the hot and cold reservoirs are not the same conserved quantity. The Raman transition, when acting adiabatically, entangles the degenerate ground spin states $\ket{\uparrow}$, $\ket{\downarrow}$ of a 3 level $\lambda$ energy system with the motional states of the quantum harmonic oscillators $\ket{n}$. With the unitary operator for this transition being
$$
\begin{aligned}
\hat{U}=&\ketbra{\uparrow}{\uparrow}\cos\left(\Omega t\sqrt{\hat{d}^\dagger \hat{d}}\right)+\ketbra{\downarrow}{\downarrow}\cos\left(\Omega t\sqrt{\hat{d}\hat{d}^\dagger}\right)\\+&i\ketbra{\uparrow}{\downarrow}\hat{d}^\dagger\frac{\sin\left(\Omega t\sqrt{\hat{d}\hat{d}^\dagger}\right)}{\sqrt{\hat{d}\hat{d}^\dagger}}+i\ketbra{\downarrow}{\uparrow}\hat{d}\frac{\sin\left(\Omega t\sqrt{\hat{d}^\dagger \hat{d}}\right)}{\sqrt{\hat{d}^\dagger \hat{d}}},
\end{aligned}
$$
where $d^\dagger d\ket{n}\approx n\ket{n}$, and $\Omega$ being the Raman Rabi rate, then for the example of a single particle in the initial state 
$$
\rho=\rho_\text{spin}\otimes\rho_{\text{energy}}=\ket{\uparrow}\bra{\uparrow}\otimes\sum_{n}\frac{e^{-nh\nu/k_BT}}{Z}\ket{n}\bra{n},
$$
the evolution produces an exchange of heat between the energy and spin degrees of freedom. This exchange liberates heat energy as work by transferring entropy into the spin reservoir, which transforms the spin equivalent of work, spin labor $\mathcal{L}$, into the spin equivalent of heat, spin therm $\mathcal{Q}$. Details of the Raman transition can be found in \cite{writeQHEoperating2018, andre2025QSHE}. Applying this unitary operation to an ensemble of particles, we can begin to realize the extent of possible work extraction. In Fig.\ref{fig: Extensive entropy storage} we show a graphical representation of the entropy transfer process as described above.

We would like to note that the adiabatic Raman transition, which acts as a heat engine, incurs a non-energetic resource cost. Specifically, the engine requires coherence to convert heat into work, as required by the second law of generalized thermodynamics \eqref{eq:Second law}. As mentioned previously, this coherence is provided by the spin labor contained within the spin polarized ensemble, $\mathcal{L}$. Therefore, the entropy battery can be viewed as a device that exploits the coherence within general quantum resources, in this case spin angular momentum, to enhance work extraction efficiency.

\subsection{Specifics of the entropy battery}

\begin{figure}[!ht]
    \centering
    \includegraphics[width=0.9\linewidth]{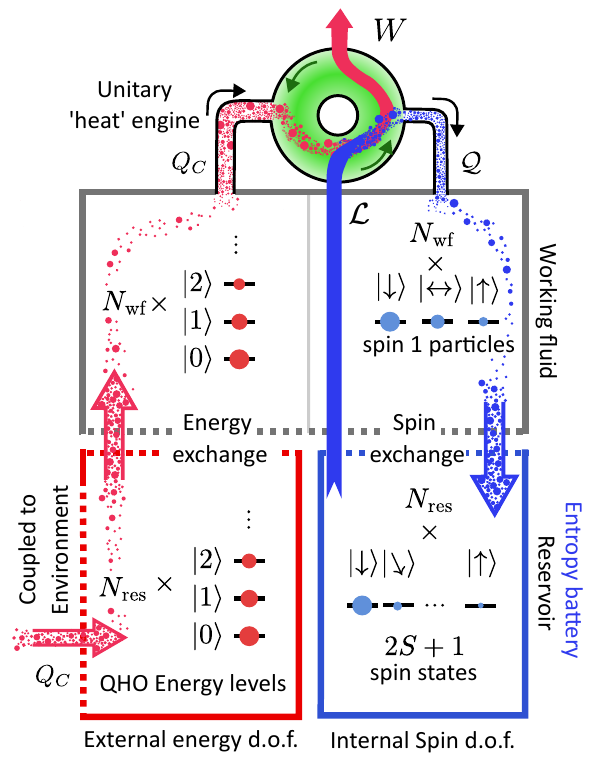}
    \caption{Schematic of the internal structure of the entropy battery. The device consists of two thermal, energy- and spin-coupled canonical ensembles: the working fluid and the reservoir, containing $N_\text{wf}$ and $N_\text{res}$ particles respectively. Each ensemble has two mutually independent conserved quantities: the external energy modes of the quantum harmonic oscillators, labeled in the number basis $\ket{n}$, and internal spin states, labeled using the arrows $\ket{\uparrow}$. The particle distributions across these states are illustrated with circles and horizontal lines, where circle size represents the population at the respective temperatures. Since $N_\text{res} \gg N_\text{wf}$, the two ensembles equilibrate rapidly in both spin and energy, so that $T_\text{res} \approx T_\text{wf}$ and $\tau_\text{res} \approx \tau_\text{wf}$. The reservoir also couples thermally to an external hot environment as in Fig.\ref{fig: Extensive entropy storage}, represented here by a canonical coupling in the bottom left. Particles in the reservoir have spin $S$, which sets the entropy capacity of the battery according to eqs.\eqref{eq:Distinguishable entropy capacity}–\eqref{eq:Fermionic entropy capacity}. In this schematic, incoherent properties (entropy) are represented with a spotted pattern, while coherent properties (information) are indicated by solid colors, illustrating the flow of entropy and information.}
    \label{fig:Entropy battery specific}
\end{figure}

The specific design of the entropy battery is shown in Fig.\ref{fig:Entropy battery specific}. The battery contains two overlapping ensembles, one small ensemble that acts as the working fluid (WF) for the unitary heat engine (adiabatic Raman transition), and the other being a much larger reservoir (RES) of spin-polarized, energetically cold particles.

The working fluid consists of $N_\text{wf}$ spin-1 particles, and the reservoir contains $N_\text{res}$ spin-$S$ particles. These ensembles continuously thermalize in both spin and energy degrees of freedom via spin and energy exchange collisions, such that at any moment in time $T_\text{wf} \approx T_\text{res}\equiv T_C$ and $\tau_\text{wf} \approx \tau_\text{res}\equiv \tau_s$. Since $N_\text{res} \gg N_\text{wf}$, we approximate the total entropy of the entropy battery as arising primarily from the reservoir, $\mathbb{S}_\text{battery} = \mathbb{S}_\text{wf} + \mathbb{S}_\text{res} \approx \mathbb{S}_\text{res}$. Once the entropy battery couples to an energetic thermal environment with unitless temperature $\tau_H\equiv T_H k_B/h\nu$, as shown in Fig.\ref{fig: Extensive entropy storage}, then the joint environment-entropy battery system can begin thermalizing. The heat engine mediated coupling allows for the transfer of entropy between the energy and spin conserved quantities across the environment and battery ensembles, under the condition $\Delta\mathbb{S}=0$, therefore

\begin{equation}
\int_{\tau_{C}}^{\tau_f}\frac{C_V}{\tau}d\tau+\int_{\tau_{s}}^{\tau_f}\frac{C_s}{\tau}d\tau = -\Delta \mathbb{S}_\text{env}(\tau_H, \tau_f).\label{eq: Reversible entropy of entropy battery}
\end{equation}
$\tau_C$ and $\tau_s$ are the battery's unitless energy and spin initial temperatures, where $\tau_C$ can be related to the absolute temperature $T_C$ by $\tau_C=T_C k_B/h\nu$ (Appendix \ref{sec: Waste response for the debye and einstein models of heat capacity}), and $\tau_H>\tau_C,\tau_s$. $\tau_f$ is the unitless final temperature once equilibrium is reached, and is common across the conserved quantity reservoirs as required by the zero'th law \eqref{eq:Zero'thlaw}. The final temperature is dependent on the entropic responses $C_s/\tau$ of the environment, and the energy/spin reservoirs of the battery, where the entropic response is the entropic equivalent of heat capacity, see Appendix \ref{sec: Waste response appendix}.

\section{Results}\label{sec: Results}

As we are interested in the theoretical potential of work extraction using the entropy battery, then we should use particle statistics that maximize the battery's entropy capacity. At face value, it would appear as though distinguishable particles would provide the highest entropy capacity according to eq.\eqref{eq:Distinguishable entropy capacity}, however, as will be discussed in \S \ref{sec: Maximising entropy capcity}, distinguishability implies non-interaction, making distinguishable particles unrealistic as our model (and real systems) require interactions between the canonical ensembles for entropy to transfer. The next best case is an ensemble of bosons. If both the environment and the battery have a large number of bosons, on the order of Avogadro's number, then the waste responses $C_s, C_V$ in the integrals of eq.\eqref{eq: Reversible entropy of entropy battery} are approximately eq.\eqref{eq:Bosonicspecificheat}, and the indefinite integrals $\int C_s/\tau d\tau$ evaluate to

\begin{equation}
\begin{aligned}
    \mathbb{S}_\text{Boson}=&\sum_{j=1}^{d-1}\left(\frac{j}{2\tau}\frac{1}{\tanh(j/2\tau)}-\ln(\sinh(j/2\tau))\right)\\&-(d-1)\ln(2).\label{eq: Boson analytic entropy}
\end{aligned}
\end{equation}
We use this expression to numerically solve for $\tau_f$ by minimising $|\Delta\mathbb{S}|$ in eq.\eqref{eq: Reversible entropy of entropy battery}, although $\tau_f$ in the thermodynamic limit (large $N$ and large initial/final temperatures) does approximately follow a geometric mean $\tau_f\approx\sqrt[3]{\tau_H\tau_C\tau_s}$, consistent with the final temperature derived by a trio of reservoirs of unequal temperatures after reaching equilibrium \cite{10.1119/1.2397094}. See Appendix \ref{sec: Endoreversible efficiency proof} for the proof.

The numerically calculated final temperature is used in the analytic boson heat expression \eqref{eq:Bosonic heat} to calculate the energetic heat transferred from the environment to the energy bath of the battery, and therefore the energetic work liberated. This represents the unitless work that is made available during a conventional heat engine cycle
$$
W_\text{conventional}=|Q_H|-|Q_C|.
$$
As discussed in the previous sections, the spin reservoir also absorbs entropy, mediated by the Raman heat engine. The Raman heat engine simultaneously liberates heat energy in the battery as energetic work at the cost of spin labor. Therefore, the total unitless energetic work of the entropy battery becomes
\begin{equation}
W_{\text{battery}} = |Q_H|-|Q_C|+|\mathcal{Q}_s|,\label{eq: Energetic work for battery}
\end{equation}
where $\mathcal{Q}_s$ is the spin therm absorbed by the battery's spin reservoir. These heats and spin equivalents are all dependent on their respective waste responses $C_s$, and so on the number of states per particle $d$ in each reservoir. The thermal energy reservoir of both the environment and the battery have in principal an infinite number of states as they are thermal motional states, however, the entropy associated with each state follows a Boltzmann-like distribution due to the nature of the thermal state \eqref{eq: Boson probability distribution over k}, with an exponential reduction as $\ket{n}$ increases. We therefore use a truncated energy state space ($d_E=400$) in our analysis without impacting accuracy. 

\begin{figure}
    \centering
    \includegraphics[width=\linewidth]{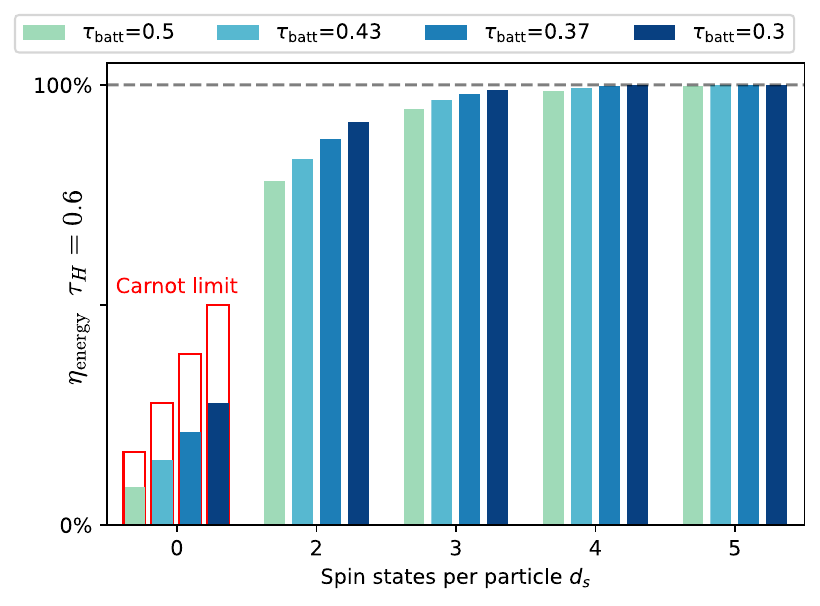}
    \caption{Efficiency of energetic work extraction $\eta_\text{energy}$ for varying number of spin states per particle calculated using the entropy and heat equations \eqref{eq: Boson analytic entropy} and \eqref{eq:Bosonic heat}. In these simulations we used $d_\text{env}=d_{\text{batt},E}=400$ without loss in accuracy. Also shown are 4 sets of initial temperatures for the ensembles energy and spin baths, with $\tau_\text{batt}=\tau_C=\tau_s$, while the environments initial temperature was set to $\tau_H=0.6$ for all cases. These unitless temperatures were chosen as it would correspond to a temperature range between room temperature, 300 K, and 600 K with a Debye cut off frequency of $21$ THz, which is a typical maximum frequency for metals \cite{rogers2005einstein}. These initial temperatures also correspond to the spin reservoir having a polarization fidelity of roughly 99.97\%. 300-600 K is a typical temperature range for coal and nuclear power plants, and concentrated solar farms. As the number of spin states increases, the energy efficiency approaches 100\%, with all initial conditions reaching approximately this limit at 5 states. This is well above the Carnot efficiency limit $\eta_\text{Carnot}=1-\tau_C/\tau_H$ which is shown by the red bars when $d=0$, with values between 16\%-50\% for the varying initial ensemble temperatures.}
    \label{fig:Energy efficiency}
\end{figure}

The energy efficiency of the battery is
\begin{equation}
\eta_\text{energy}=\frac{W_\text{battery}}{|Q_H|}=1-\frac{|Q_C|-|\mathcal{Q}_s|}{|Q_H|},\label{eq: Energy efficiency of battery}
\end{equation}
with the results shown in Fig.\ref{fig:Energy efficiency} for varying initial battery temperatures. Our results show that once $d_s\ge2$, the energy efficiency surpasses the Carnot limit. At 5 spin states, all initial temperatures considered reach close to $100\%$ energy efficiency. This occurs despite the probability distributions in each degree of freedom being initially states of maximum entropy (generalized Gibbs distribution), without induced quantum coherence. Further, as the entropy transfer occurs adiabatically, the heat engines are all endoreversible. The energy efficiency with 0 spin states follows $1-\sqrt{\tau_C/\tau_H}$, making our model consistent with the expected efficiency one would obtain when running the entropy battery at maximum power \cite{NOVIKOV1958125, 10.1119/1.10023, 10.1119/1.2397094}. See Appendix \ref{sec: Endoreversible efficiency proof} for further details. 

\section{Discussion}\label{sec: Discussion}

The Carnot limit sets the maximum efficiency achievable by a heat engine operating between two thermal energy reservoirs of different temperatures. As mentioned in \S \ref{sec: The application of quantum information}, a common strategy in quantum heat engines to surpass this limit is to induce quantum coherence within the thermal bath, either via squeezed thermal states \cite{PhysRevX.7.031044, PhysRevE.86.051105, PhysRevE.93.052120} or other induced coherences \cite{correa2014quantum, scully2003extracting, dillenschneider2009energetics}. All of these devices operate below the generalized Carnot limit \cite{Abah_2014}, therefore there is no real quantum advantage. In contrast, the entropy battery achieves super-Carnot efficiencies in large ensembles using only maximum entropy states, characterised by their unitless temperatures, demonstrating that this limit can be surpassed only when an energetically hot source is coupled to an ensemble with multiple conserved quantities via multiple unitary heat engines.

It must be re-iterated that this increase in energy efficiency comes at the cost of spin labor, with the generalized work of the battery being
$$
\mathbb{W}=|Q_H| - |Q_C| + |\mathcal{Q}_s| - |\mathcal{L}|,
$$
where $|\mathcal{L}|=|\mathcal{Q}_s|$. The generalized efficiency of the battery when considering all resource costs therefore returns to the Carnot efficiency limit $\eta=1-|Q_C|/|Q_H|$, which proves that the entropy battery does not gain a `free-lunch'. Eq.\eqref{eq: Energy efficiency of battery} is therefore the \textit{energetic} efficiency of the battery. If the initial preparation of spin labor within the battery requires energy, and if this energy cost is equal to the energy saved during the operation of the entropy battery, then the entropy battery would not have an advantageous net energy efficiency. However, re-polarising spin reservoirs does not necessarily require an energy input, as naturally occurring sources of spin coherence can instead be used to re-polarize the reservoir. Consequently, the entropy battery now has an additional advantage: it extends the range of usable natural resources for energetic work extraction beyond energy alone. 

The unitary operation that couples the external energy and internal spin degrees of freedom is also controllable. Combining this controllability with the assumed mutual independence of the spin degree of freedom, the state of `charge' of the battery would remain approximately constant even in an energetically hot environment. This makes the entropy battery useful in applications that require such controllability, with it effectively performing the job of a conventional battery, all the while storing not energy, but information in the form of coherence when charged. With this coherence stored across multiple conserved quantities but within the same $N_{\text{res}}$ particles, then a battery of this kind could in principal extract more energetic work per particle than storable by conventional batteries.

Similarly, the entropy battery can function as a heat storage device. When the environment-battery system thermalizes, the final unitless temperature $\tau_f$ is lower than the initial unitless temperature of the hot environment ($\tau_f < \tau_H$), meaning that the entropy battery absorbs and stores heat. Because the stored heat is distributed between the other non-energetic conserved quantities, the entropy battery achieves a higher density of heat storage than a single cold thermal bath. Such a mechanism could find applications in systems requiring efficient heat removal, such as spacecraft operating in vacuum or high-performance computing.

This article focuses on an ideal scenario in which the battery-environment systems thermalize adiabatically. We briefly note the role of open-system effects here. Irreversibility arising from an injection of entropy from an external source, or non-independent conserved quantities, such that the total entropy is non-extensive, $\mathbb{S}_\text{battery} < \mathbb{S}_\text{energy} + \mathbb{S}_\text{spin}$ both effect the effective entropy capacity of the battery. Both of these effects increases the final unitless temperature once general equilibrium is reached, and decreases the extractable work. Beyond improving isolation or enhancing the mutual independence of the conserved quantities, one could compensate for this loss by increasing the number of states, however that analysis is beyond the scope of this article.

A particular field that may apply the concepts developed in this article is quantum error correction. The decoherence of a qubit over time can be described as a thermalization process with a thermal environment \cite{PhysRevLett.108.080402,PrathikCherian_2019}, and as such, one can imagine that having access to another conserved quantity reservoir that is independent of the basis of measurement (the qubit) within a single physical system (the ion, superconductor etc.) would allow for the `re-direction' of decoherence into the unused degree of freedom. If possible, this would slow the decoherence time of the measured qubit, potentially leading to less dependence on error correcting qubits. Actual implementation of such an error correcting technique is also beyond the scope of this article, however NMR algorithmic cooling \cite{doi:10.1073/pnas.241641898,fernandez2004algorithmic} bears some similarities in terms of using ancillary qubits (the spin states) for reversibly cooling the measurement qubit (the energy states).

\section{Maximizing entropy capacity}\label{sec: Maximising entropy capcity}
The analysis presented in this article focuses on a single operating cycle, after which the environment ensemble is left cooler than its initial state, $\tau_f < \tau_H$. Repeated operation of the battery, assuming the environment is reset to its original temperature after each cycle, would allow continuous extraction of energetic work until the battery reaches the same unitless temperature as the now effectively isothermal environment. These extra cycles increase the amount of work extractable from the hot thermal energy bath. Therefore, it is of interest to understand how one could maximize the battery's entropy capacity. According to eqs.\eqref{eq:Distinguishable entropy capacity} and \eqref{eq:Bosonic entropy capacity}, one could increase the number of particles $N$ or states $d$, as either of these increases the number of microstates $\Omega$ and therefore the entropy capacity. Another way would be to increase the number of independent conserved quantities. The dependence of the entropy capacity on these variables can be understood directly by application of the maximum entropy principle \cite{jaynes1957information}.

Consider a particle confined within a closed volume and isolated from any external interactions. Let the particle possess a total of $M$ conserved quantities, comprising both external (e.g., kinetic energy, momentum) and internal (e.g., spin, charge) degrees of freedom. For the $m^\text{th}$ unitless conserved quantity $A_m$, there are $d_m$ orthogonal eigenstates with eigenvalues $j$, labeled in the computational basis $j=0,1,2\dots, d_m-1$.  Expanding the system to $N$ particles, we now introduce the definition of the macrostate, denoted $a_j$. Each particle can occupy a discrete state $j$, where there are $k_j$ such particles in the same state. The macrostate is then defined as $a_j=jk_j$, where $\sum_j k_j=N$. For any macrostate there can be, in principal, many microstate. We denote these microstates $\Omega_m$, with $\Omega_m$ being the set of \textit{distinguishable} configurations of particles across the $j$ states for the $m^\text{th}$ conserved quantity. In the most general case, permutations of individual particles add a combinatorial factor to the number of microstates that share the macrostate $a_j$. For example, if the particles are distinguishable, then there are $N!/(k_0!k_1!...)$ microstates for each $a_j$. In contrast, if the particles are indistinguishable then there are no such distinguishable permutations. Importantly, the Shannon entropy must be calculated as a sum over all distinguishable microstates
\begin{equation}
\mathbb{S}=-\sum_{\Omega} P(\Omega)\ln P(\Omega),\label{eq: Shannon entropy}
\end{equation}
where $\Omega$ is the set of microstates across every conserved quantity. 


The probability distribution that maximizes entropy, under the constraints of probability normalization $\sum_{\Omega} P(\Omega)=1$, Shannon entropy \eqref{eq: Shannon entropy}, and the $M$ total conserved quantities
\begin{equation}
\begin{aligned}
\braket{A_m}=&\sum_{\Omega} \left(\sum_{j=0}^{d-1}a_j(\Omega)\right)P(\Omega)\\=&\sum_{\Omega} a(\Omega)P(\Omega),
\end{aligned}
\end{equation}
is \cite{jaynes1957information}
\begin{equation}
P(\Omega)=\prod_{m=0}^Mc_m P_m(\Omega, \gamma_m)=\prod_{m=0}^{M}c_m e^{-\gamma_m a_m(\Omega)},\label{eq: Separable conserved quantity probability distribution}
\end{equation}
where $\gamma_m=\tau_m^{-1}$ and $c_m$ are the Lagrange multiplier and normalization constant for the $m$\textsuperscript{th} conserved quantity. In this article, all ensembles are treated as canonical, with no particle exchange and therefore no associated chemical potential. The ensemble normalization $\prod_m c_m=Z$ is,
$$
Z=\sum_\Omega \prod_{m=0}^M e^{-\gamma_m a_m(\Omega)},
$$
where $Z$ is the partition function. The entropy is therefore generally \cite{jaynes1957information}
$$
\mathbb{S}=\ln(Z)+\sum_{m=0}^M\gamma_m\braket{A_m}.
$$
As defined, this entropy is sub-additive over the conserved quantities $\mathbb{S}\le \sum_{m} \mathbb{S}_m$ as $Z$ is not necessarily separable over $m$, $Z\neq Z_1Z_2\dots Z_M$. However, if the conserved quantities are assumed to be mutually independent, then the total partition function factorises as $Z=\prod_{m}Z_m$, and the entropy becomes extensive, $\mathbb{S}=\sum_m \mathbb{S}_m$ where
$$
\mathbb{S}_m=\ln(Z_m)+\gamma_m\braket{A_m}.
$$
For the conserved quantity $A_m$, the normalized probability distribution is
\begin{equation}
P(\Omega_m,\gamma_m)=\frac{e^{-\gamma_m a_m(\Omega_m)}}{Z_m}\label{eq: General probaiblity distribution}
\end{equation}
where
$$
Z_m=\sum_{\Omega_m}e^{-\gamma_m a_m(\Omega_m)}.
$$
The microstates $\Omega_m$ of the $m$\textsuperscript{th} conserved quantity are now independent of the other conserved quantities. Therefore, mutual independence over the conserved quantities maximizes the ensembles entropy capacity. To further increase entropy capacity, we are limited to increasing the number of microstates associated with the 
$M$ conserved quantities.

\subsection{Distinguishability}\label{sec: Distinguishability subsection}
\begin{figure}
    \centering
    \includegraphics[width=\linewidth]{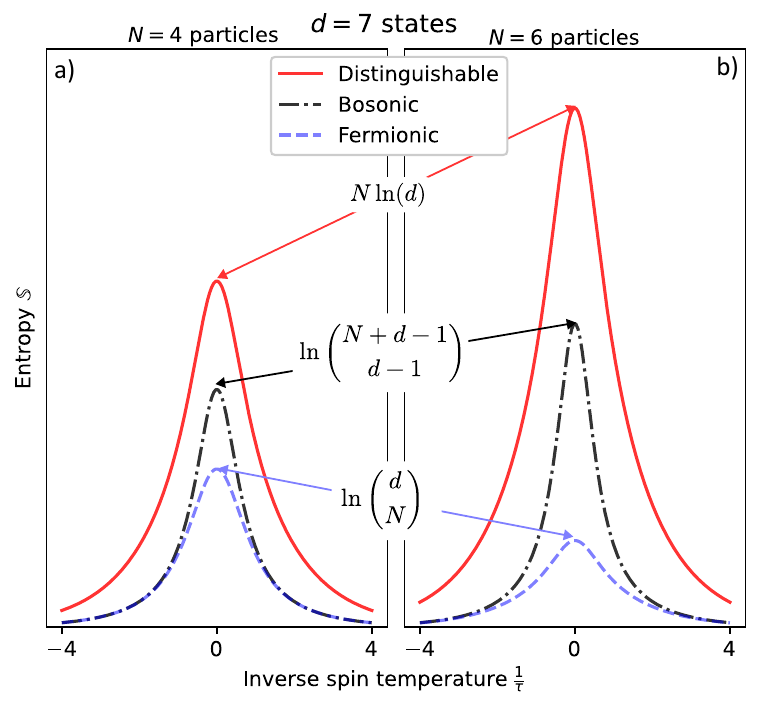}
    \caption{An example entropy distribution over spin temperature for the different particle statistics, distinguishable, bosonic and fermionic with 7 states. In (a) and (b) we show example distributions with $N=4$ and $N=6$ particles respectively, illustrating how distinguishable and bosonic maximum entropies increase with $N$ without bound, while the fermionic maximum entropy decreases as saturation is approached $N=d$ where $\mathbb{S}_\text{Fermi}=0$ for all temperature. Maximum entropy, or the systems entropy capacity is given by the entropy at infinite temperature $\tau=\infty$, which we have labeled for the different ensemble statistics.}
    \label{fig: Comparing distributions}
\end{figure}
As shown in Fig.\ref{fig: Comparing distributions}, an ensemble of distinguishable particles achieves the largest entropy capacity. The partition function of such an ensemble is
$$
Z_\text{dist}=\sum_\mathbf{k}\frac{N!}{k_0!k_1!...k_{d-1}!}\prod_{j=0}^{d-1}e^{-\gamma  a_j},
$$
where $\mathbf{k}=\{k_0,k_1,...,k_{d-1}\}$ specifies a configuration of the $k_j$ particles over the $j$ states. See Appendix \ref{sec: Distinguishable particles} for further details. This partition function is immediately factorizable for every one of the $N$ particles, since it is identically a multinomial expansion
$$
Z_\text{dist}=\left(\sum_{j=0}^{d-1}e^{-\gamma j}\right)^N=Z^N.
$$
This separability applies to the probability distribution \eqref{eq: Seperability of the probability distribution}
$$
P(\Omega, \gamma)=\left(\frac{e^{-\gamma a(\Omega)}}{Z}\right)^N=p(\gamma)^N,
$$
where $p(\gamma)$ is the single particle Boltzmann distribution. The entropy at infinite temperature $\gamma=1/\tau=0$ is therefore proportional to the number of particles
$$
\mathbb{S}_\text{max}=N\ln(d).
$$
A separable probability distribution and extensive entropy can only occur if the particles are mutually independent. In the distinguishable particle ensemble, this mutual independence holds for each of the $N$ particles. Mutual independence therefore implies non-interaction, where interaction is defined generally as an event in which particles exchange a conserved quantity under the constraint of the relevant conservation law, or equivalently as a transfer of mutual information between particles. Consequently, distinguishable particles that maintain distinguishability at all times must be non-interacting, an argument similarly made by Jaynes in \cite{jaynes1965gibbs}. This has important implications for the entropy capacity of a realizable entropy battery. For unitary heat engines to operate, or for two canonical ensembles to thermalize, interactions between particles are required. For these reasons, distinguishable particles cannot be used in the entropy battery. Further discussion can be found in Appendix \ref{sec: Distinguishability}.

The next best scenario is an ensemble of bosons. As discussed in Appendix \ref{sec: boson case}, unlike distinguishable particles, the probability distribution for an ensemble of bosons is inseparable over each of the particles. Therefore, as a contrasting argument to the distinguishable particles, bosons \textit{must} allow interaction, such that one cannot describe the particles as individuals after there becomes an exchange of information. The description we give implies that it is the mutual information upon interaction that ultimately determines a particles distinguishability. If the information is mutually exclusive then there remains a form of distinguishability, yet no interaction, but if the information is mutually dependent then we lose distinguishability and there must have been an interaction. In this sense, what is ultimately indistinguishable upon interactions is the information that is encoded in the conserved quantities of the system. This is especially apparent when monitoring the mutual exchange of information between typically mutually independent conserved quantities like that used in the entropy battery. The exchange of generalised heat and work between conserved quantities is best understood under the context of information transfer. 

Therefore, a system of bosons is the best case scenario for our model of the entropy battery, as they maximize entropy capacity while allowing information and entropy to be transferred between both the canonical ensembles of the battery and environment, and the independent conserved quantity.

\section{Conclusion}\label{sec: Conclusion}
From the perspective of information thermodynamics, an ensemble of particles with multiple, mutually independent conserved quantities exhibits an enlarged entropy capacity compared to classical, energy only canonical ensembles. By applying unitary operations that couple these conserved quantities, this entropy capacity can be harnessed during work extraction, yielding energy efficiencies beyond the Carnot limit. We refer to the ensemble as being an entropy battery, with it storing entropy as it thermalizes with an external reservoir. We justify treating spin angular momentum as an independent thermodynamic system by showing that, like energy, spin reservoirs obey the laws of generalized thermodynamics, follow Bose-Einstein and Fermi-Dirac statistics, satisfy the fluctuation-dissipation and equipartition theorems (see Appendix \ref{sec: Ensemble multpliicities}), and have analogous Einstein solid and Debye models of heat capacity. We also discuss the conditions under which particles are distinguishable (Appendices \ref{sec: Distinguishable particles} and \ref{sec: Distinguishability}), concluding that distinguishable particles can exist only if they are non-interacting. Consequently, the apparent entropy capacity of such particles cannot be used for energetic work extraction via heat engines. This leads naturally to a generalization of indistinguishability in informational terms, independent of the specific conserved quantities or particles, and provides a conceptual understanding of the entropy battery as a device that transfers entropy through subsystem entanglement.

\section*{Acknowledgements}
I thank the ARC Centre of Excellence (LP180100096) and the Lockheed Martin Corporation for their financial support. I also express my deepest appreciation to my mentors, Assoc. Prof. Erik W. Streed, Prof. Joan Vaccaro, and Dr. Mark Baker, for their guidance and insightful feedback throughout the development of this work.

\section*{Data availability statement}
The data that support the findings of this article are openly available \cite{McClelland_entropy-battery_2025}.

\providecommand{\noopsort}[1]{}\providecommand{\singleletter}[1]{#1}%
%


\appendix
\section{Information Thermodynamics}\label{sec: Spin temperature}
In this section we derive, and present where assumed, the laws of spin thermodynamics, with these laws applying equally to arbitrary mutually independent conserved quantity. 

From the probability distribution in eq.\eqref{eq: General probaiblity distribution}, the Shannon entropy \eqref{eq: Shannon entropy} of the spin component of the ensemble when mutually independent of any other degree of freedom is
\begin{equation}
\mathbb{S}=\ln(Z)+\gamma\braket{\bar{J}_z},
    \label{eq: entropy}
\end{equation}
where $\braket{\bar{J_z}}=\braket{J_z}/\hbar$ and $\gamma$ are the unitless average angular momentum and spin Lagrange multipliers respectively. We have removed the `$m$' label for brevity. From eq.\eqref{eq: entropy}, we show a familiar, and generally known relation
\begin{equation}
    \frac{\partial\mathbb{S}}{\partial \braket{\bar{J}_z}}=\gamma\label{eq: Entropy momentum relation}
\end{equation}
\cite{jaynes1957information2}, which is directly analogous to the thermodynamic relation of energy and entropy \cite{schrödinger1948statistical}
\begin{equation}
    \frac{\partial \mathbb{S}}{\partial \braket{E}}=\frac{1}{T}.\label{eq: entropy energy relation}
\end{equation} 
This leads to the interpretation of the Lagrange multiplier of the average spin angular momentum constraint, $\gamma$, as the inverse spin temperature $\tau=\frac{1}{\gamma}$ \cite{Guryanova_2016}. From eq.\eqref{eq: General probaiblity distribution}, we also obtain expressions of average angular momentum in terms of the partition function
\begin{equation}
\braket{\bar{J}_z}=\tau^2\frac{\partial}{\partial \tau}\ln(Z)=-\frac{\partial}{\partial\gamma}\ln(Z)\label{eq: Angular momentum in terms of partition function},
\end{equation}
which is analogous to the expression of average energy
$$
\braket{E}=-\frac{\partial}{\partial\beta}\ln(Z)
$$
where $\beta=k_BT$ is the energy Lagrange multiplier \cite{jaynes1957information,jaynes1957information2}. Negative spin temperatures appear due to having finite eigenstates, which is a phenomena also found in energetic systems with finite energy eigenstates \cite{Braun_2013, PhysRev.103.20}.

Showing the zero'th and third laws of spin thermodynamics is straightforward from here. Consider a spin ensemble with many particles, that is also in a state of maximum entropy such that it's probability distribution is described by eq.\eqref{eq: General probaiblity distribution} with spin temperature $\tau_1$. Take another similar spin system that is independently at a state of maximum entropy, with spin temperature $\tau_2$, where initially $\tau_1\neq\tau_2$. Upon spin exchange contact without particle exchange, the ensembles evolve in what can be considered a spin thermalization process. By the first law of generalized thermodynamics, the net change in spin between the ensembles is 0
\begin{equation}
\delta \braket{\bar{J}_{z,1}}=-\delta \braket{\bar{J}_{z,2}}\label{eq: First law of spin thermodynamics}.
\end{equation}
Since this is a closed system, then the net change in entropy during thermalization must also be 0,
\begin{align}
    \delta \mathbb{S}_1&=-\delta \mathbb{S}_2\label{eq: Second law}.
\end{align}
Eq.\eqref{eq: Second law} would imply that in principal the system is reversible as it follows Liouvillian trajectories, which are deterministic. Eq.\eqref{eq: Second law} is the second law of spin thermodynamics for freely evolving isolated systems.

As these two ensembles are initially independent, then both the total spin angular momentum and total entropy of each ensemble add extensively. Changes in a systems spin angular momentum therefore has a corresponding change in entropy, and this is balanced by an equivalent change in the other system. During the spin thermalization process these systems will exchange spin and entropy deterministically up until the ratio $\delta J_z/\delta S$ for the two systems are equal, which comes from the division of eqs.\eqref{eq: First law of spin thermodynamics} and \eqref{eq: Second law}
$$
\begin{aligned}
    \frac{\delta \braket{\bar{J}_{z,1}}}{\delta\mathbb{S}_1}=\frac{\delta \braket{\bar{J}_{z,2}}}{\delta \mathbb{S}_2}.
\end{aligned}
$$
For ensembles with large $N$, the fluctuations of spin angular momentum and entropy between the baths become vanishingly small with respect to the average equilibrium values, as the relative variation scales with $1/\sqrt{N}$ \cite{MARCONI_2008}. We can therefore approximate the small changes as an infinitesimal, and make use of the result in eq.\eqref{eq: Entropy momentum relation} to show that this can only be true if

\begin{equation}
\tau_1=\tau_2,\label{eq: Zero'th law}
\end{equation}
meaning the two systems will reach spin exchange equilibrium when $\tau_1=\tau_2$. This is the zero'th law of spin thermodynamics.

Finally, the entropy of this system will approach 0 as the spin temperature approaches 0. This can be shown explicitly using the expression of the probability distribution \eqref{eq: General probaiblity distribution}. As $\tau\rightarrow 0$
$$
P(\Omega)=\begin{cases}0,& k_0\text{ or }k_{d-1}\neq N\\1, &k_0\text{ or }k_{d-1}=N,\end{cases}
$$
which is to say the spin system becomes maximally polarized. In this case, the entropy \eqref{eq: entropy} is 0 for all microstates, and so the total entropy is 0. This is the third law of spin thermodynamics.


Aside from the laws of thermodynamics applying to spin angular momentum, Vaccaro et. al. have explored the concepts of spin `heat' and spin `work' , renamed spin therm $\mathcal{Q}$ and spin labor $\mathcal{L}$ respectively \cite{manakil2025GSHE,writeQHEoperating2018}. It was shown that heat engines operating with spin systems can be written in terms of these analogous thermodynamic properties.

We conclude that spin, when treated as an independent conserved quantity, follows the same laws of thermodynamics as energy. It can be expected that any application of thermodynamics, such as heat engines, would also work within spin systems. In the case of spin heat engines that operate between two spin reservoirs, the `work' that is extracted will be in the form of coherent spin angular momentum, or spin labor $\mathcal{L}$. However, unitary coupling of the energy basis to the spin basis within a single system can allow for the extraction of energetic work \cite{manakil2025GSHE, writeQHEoperating2018}. As spin follows the laws of thermodynamics, then thermodynamics is clearly not a theory of only energy and heat. With this section applying to arbitrary unitless mutually independent conserved quantity $A_m$ by substitution, then a general interpretation of these results is that thermodynamics is fundamentally an information theory.

\section{Ensemble Probability distributions}
\label{sec: Ensemble multpliicities}
\subsection{Distinguishable particles}
\label{sec: Distinguishable particles}
 An ensemble of labeled particles are, by definition, distinguishable. The exact labeling scheme one might use is discussed in Appendix \ref{sec: Distinguishability}. For this ensemble, the number of microstates $\Omega$ increases as permutations of the particles that are in a particular configuration $\mathbf{k}=\{k_0, k_1,... k_{d-1}\}$ constitute new microstates. The multiplicity of state configurations $g(\mathbf{k})$ that result in the same macrostate $a(\Omega)$ if all particles are distinguishable from one another is a multinomial
$$
g_\text{dist}(\mathbf{k})=\begin{pmatrix}N\\k_0,k_1,...k_{d-1}\end{pmatrix}=\begin{pmatrix}N\\\mathbf{k}\end{pmatrix}.
$$
This multiplicity is the same used by Bernoulli and Laplace for distinguishable particles \cite{janes1982rationale}. The partition function can then be represented as
$$
\begin{aligned}
Z_\text{dist}=
 {\sum_{\mathbf{k}}} \begin{pmatrix}N\\\mathbf{k}\end{pmatrix}\prod _{j=0}^{d-1} e^{-\gamma m_jk_j}\\
\end{aligned}
$$
which is a multinomial expansion of the variables $e^{-\gamma m_j}$, and so the partition function can be simplified to
\begin{equation}
Z_\text{dist}=\left(\sum_{j=0}^{d-1} e^{-\gamma m_j}\right)^N=Z^N\label{eq: Distinguishable partition function}.
\end{equation}
The partition function for distinguishable particles is therefore separable into a product of the per particle partition function $Z=\sum_{j=0}^{d-1} e^{-\gamma m_j}$. This separability applies also to the probability distribution. From eq.\eqref{eq: General probaiblity distribution},

\begin{align}
P(\Omega)&=\frac{e^{-\gamma\sum_{j=0}^{d-1}m_jk_j}}{Z^N}\notag\\
P(\Omega)&=\left(\frac{e^{-\gamma m_0 k_0}}{Z^{k_0}}\right)\left(\frac{e^{-\gamma m_1k_1}}{Z^{k_1}}\right)\dots\notag\\
&=\prod_{j=0}^{d-1}p(j)^{k_j}=p(j)^N\label{eq: Seperability of the probability distribution}
\end{align}
where $p(j)=e^{-\gamma m_j}/Z$ is the Boltzmann distribution for a single particle in state $j$. Eq.\eqref{eq: Seperability of the probability distribution} is separable down to the individual particles.

The separability of the particles’ probability distributions is significant, as probability theory requires that such particles be mutually independent. Each particle must act as an independent entity, unable to influence the outcomes of the others. Consequently, particles can only be considered distinguishable if they are non-interacting. This is discussed further in Appendix \ref{sec: Distinguishability}.


\subsubsection{Polarization, spin temperature relation}\label{sec: Polarisation, gamma relation}
\begin{figure}[b]
    \centering
    \includegraphics[width=\linewidth]{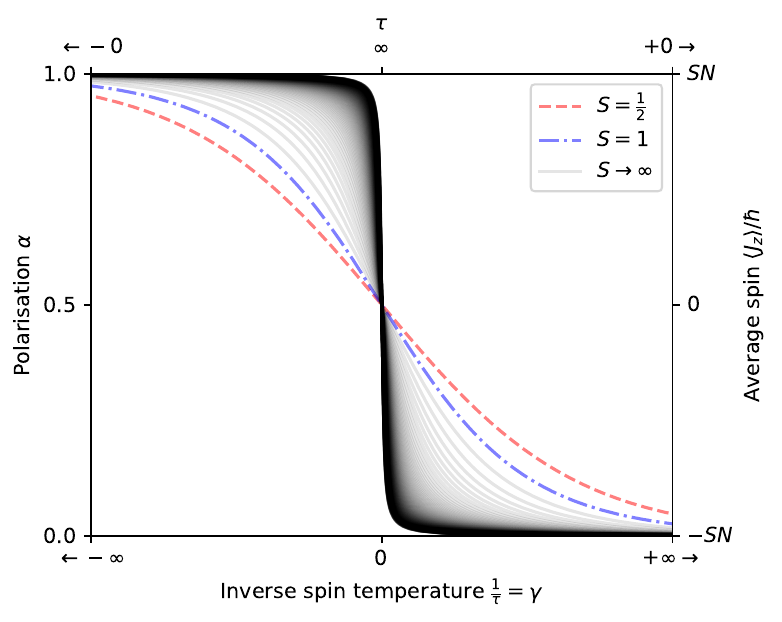}
    \caption{Mapping of spin temperature $\tau$ to polarization, eq.\eqref{eq: S=1/2 tau to polarisation}, for different ensembles of distinguishable particles with per particle spins $S$. This plot shows up to $S=200$, but trends towards a steepening slope around $\gamma=0$, where in the limit of $S\rightarrow\infty$ one would expect a step function centred at $\gamma=0$}
    \label{fig:Polarisation to gamma relation}
\end{figure}
It is convenient to re-express the spin temperature $\tau$ in terms of the spin-polarization $\alpha$, which provides a mapping to the macroscopic observable $J_z$. From eq.\eqref{eq: Angular momentum in terms of partition function}

$$
\begin{aligned}
    \left(2\alpha S-S\right)N&=-\frac{\partial}{\partial\gamma}\ln\left(Z_\text{dist}\right)\\
    2\alpha S-S&=-\frac{\partial}{\partial\gamma}\ln\left(\sum_{j=0}^{d-1}e^{-\gamma m_j}\right)\\
\end{aligned}
$$
where $\braket{\bar{J}_z}=(2\alpha-1)SN$. From this we obtain a polynomial

\begin{align}
     2\alpha S-S&=\frac{\sum_{j=0}^{d-1}m_je^{-\gamma m_j}}{\sum_{i=0}^{d-1}e^{-\gamma m_j}}\notag\\
     0&=\sum_{j=0}^{d-1}e^{-\gamma m_j}(2\alpha S-m_j-S)\label{eq: Root solution to gamma, alpha relation}.
\end{align}
Since spin angular momentum is quantised with evenly spaced spin states, then $m_j=j-S$ and eq.\eqref{eq: Root solution to gamma, alpha relation} can be written as
$$
0=\sum_{j=0}^{d-1}e^{-\gamma j}(2\alpha S-j).
$$
This polynomial is of the $d-1=2S$ order for the variable $e^{-\gamma}$ with coefficients $2\alpha S-j$. Solving the root provides the mapping of spin temperature to polarization
$$
\tau=\ln\left(\frac{1}{f(\alpha)}\right)^{-1},
$$
where $f(\alpha)$ is the real positive root of the polynomial $\sum_{j=0}^{d-1}x^j(2\alpha-j)$. The roots must be real and positive as $\tau$ must be real, and $\ln(x)$ requires positive arguments to compute. 

Some example solutions that can be analytically expressed include the $S=\frac{1}{2}$ (linear) and $S=1$ (quadratic) systems

\begin{equation}
\begin{aligned}
    S=\frac{1}{2}:&\quad\tau=\ln\left(\frac{1-\alpha}{\alpha}\right)^{-1}\label{eq: S=1/2 tau to polarisation}\\
    S=1:&\quad\tau=\ln\left(\frac{4(1-\alpha)}{(2\alpha-1)+\sqrt{-12\alpha^2+12\alpha+1}}\right)^{-1}
\end{aligned}
\end{equation}
Higher-order solutions are solved numerically and are presented in Fig.\ref{fig:Polarisation to gamma relation}. These relations of temperature to polarization are equivalent to expressions found in the equipartition theorem for discrete systems \cite{davis2024temperatureequipartitiondiscretesystems}, with our work generalizing to an arbitrary number of orthogonal states $d=2S+1$ for distinguishable particles.

\subsection{Bosons}\label{sec: boson case}

For a canonical ensemble of indistinguishable bosons, any permutations on the configuration of particles $\mathbf{k}$ is not itself a new microstate like that for distinguishable particles. Therefore, the multiplicity $g(\mathbf{k})=1$ for all configurations of $\mathbf{k}$, and the microstates $\Omega=\mathbf{k}$. The probability distribution from eq.\eqref{eq: General probaiblity distribution} now becomes

\begin{equation}
    P_\text{Boson}(\mathbf{k})= \frac{e^{-\gamma \sum_{j=0}^{d-1}m_j k_j}}{Z_\text{Boson}},\label{eq: Boson probability distribution over k}
\end{equation}
where 
\begin{equation}
Z_\text{Boson}=\sum_\mathbf{k}e^{-\gamma\sum_{j=0}^{d-1}m_jk_j}.\label{eq: Boson parition function in the k index}
\end{equation}
By changing the index from the particle configuration $\mathbf{k}$ to the macrostate $a(\mathbf{k})=\sum_{j=0}^{d-1} m_jk_j$, we show that spin bosons follow Bose-Einstein statistics \cite{bose1994man}. The macrostate index $a(\mathbf{k})$ ranges from $-SN$ to $SN$, which we can shift to the computational macrostate $m=a(\mathbf{k})+SN=\sum_{j=0}^{d-1}jk_j$ without affecting the distribution. The multiplicity using this index is derived in Appendix \ref{subsubsec: Multiplicity of indistinugishable particles}, and is explicitly the coefficient of the $[x^m]$\textsuperscript{th} term of a Gaussian binomial,
\begin{equation}
g_{\text{Boson}}(m)=[x^m]\begin{pmatrix}N+d-1\\d-1\end{pmatrix}_x.\label{eq: Boson multiplicity with n index}
\end{equation}
This shows that the partition function is
\begin{equation}
Z_\text{Boson}=\begin{pmatrix}N+d-1\\d-1\end{pmatrix}_p:\quad p=e^{-\gamma}.\label{eq: Boson partition function with m index}
\end{equation}
Both eqs.\eqref{eq: Boson parition function in the k index} and \eqref{eq: Boson partition function with m index} are equivalent, however they do not automatically allow for separable distributions like that for the distinguishable particles. These expressions obey Bose-Einstein statistics, with $\sum_m g_\text{Boson}={}^{N+d-1}C_{N}$ being the multiplicity used by Bose for an ensemble of bosons in the energy basis \cite{bose1994man}. Going further, we explicitly show that spin bosons follow the Bose–Einstein distribution in the limit of an infinite number of particles. The partition function \eqref{eq: Boson partition function with m index} can then be expressed as a generating function

\begin{align}
Z_\text{Boson}&=[t^N]\sum_{k_j=0}^\infty\begin{pmatrix}k_j+d-1\\d-1\end{pmatrix}_{e^{-\gamma}} t^{k_j}\\&=[t^N]\prod_{j=0}^{d-1}\frac{1}{1-te^{-\gamma m_j}},\label{eq: Newtons generalised binomial theorem}
\end{align}
where $[t^N]$ is a selection variable, used to select the polynomials of $t$ that correspond to $N$ particles, and eq.\eqref{eq: Newtons generalised binomial theorem} is an identity from Newton's generalized binomial theorem. Eq.\eqref{eq: Newtons generalised binomial theorem} demonstrates that, in the infinite-particle limit, the boson partition function becomes separable over the macrostates

\begin{equation}
Z_\text{Boson}=[t^N]\prod_{j=0}^{d-1}z_\text{Boson}(j,t)\label{eq: Seperabilty of boson partition function}.
\end{equation}
As in the distinguishable case (Appendix \ref{sec: Distinguishable particles}), this implies that in the infinite-particle limit, the probability distributions of each state $j$ are mutually independent, though the particles themselves are not. The state dependent probability distribution is
$$
P(j)=\frac{e^{-\gamma m_jk_j}}{z(j)},
$$
where 
$$
z(j)=\frac{1}{1-e^{\gamma m_j}}=\sum_{k_j=0}^\infty e^{-\gamma m_j k_j}.
$$
The selection variable $t$ was removed as the states $j$ are now effectively mutually independent grand canonical ensembles, and we are no longer considering a finite system. The average occupation of particles in state $j$ is then

\begin{align}
\braket{n_j}&=\sum_{k_j=0}^\infty k_jP(j)=\frac{1}{z(j)}\sum_{n=0}^\infty k_je^{-k_j\gamma(j-S)}\notag\\
&=\frac{1}{z(j)\gamma}\left(\frac{\partial }{\partial S}z(j)\right)\notag\\
&=\frac{1}{e^{\gamma m_j}-1},\label{eq: Bose-einstein distribution for spin}
\end{align}
where we used $m_j=j-S$. Eq.\eqref{eq: Bose-einstein distribution for spin} represents the Bose–Einstein distribution \cite{bose1994man} in the spin degree of freedom. The separability of the probability distributions occurs only in the limit of an infinite number of particles, and only separable over the states. Since this separability does not apply to individual particles, the particles themselves are not mutually independent and bosons must allow interaction.
\subsection{Fermions}\label{sec: Fermion case}
Fermionic systems modify the ensemble statistics further due to the antisymmetric nature of particle exchange, which gives rise to the Pauli exclusion principle \cite{pauli1924pauli}. We can no longer assume that an arbitrary number of particles may occupy the same spin state, as this would violate the exclusion principle if the particles also shared identical values in all other conserved quantities.

Assuming the particles occupy identical states across all other degrees of freedom, the system can contain at most $N=d$ particles, corresponding to one particle per spin state. Therefore, this system is constrained by
\begin{equation}
    \sum_{j=0}^{d-1}k_j=N\le d;\quad k_j\in\{0,1\}.
    \label{eq: Fermi particle configuration}
\end{equation}
The set of microstates $\Omega$ is then $\Omega=\mathbf{k}_\text{Fermi}=\{k_j\in\{0, 1\}:k_0, k_1,...,k_{d-1}\}$, and the probability distribution is
$$
P(\mathbf{k}_\text{Fermi})=\frac{e^{-\gamma\sum_{j=0}^{d-1}jk_j}}{Z_\text{Fermi}}.
$$
where
$$
Z_\text{Fermi} = \sum_{\mathbf{k}_\text{Fermi}}e^{-\gamma\sum_{j=0}^{d-1}jk_j}
$$

As in the bosonic case, by changing to the macrostate index $m$, we can explicitly show that spin fermions follow Fermi–Dirac statistics. The multiplicity of the macrostate $m=a(\mathbf{k}_\text{fermi})+SN=\sum_{j=0}^{d-1}jk_j$ is calculated using the generating function
$$
g_{\text{Fermi}}(m)=[t^Nx^m]\prod_{j=0}^{d-1}(1+tx^j)
$$
where $[t^N x^m]$ selects the polynomial terms that correspondingly have $N$ atoms with macrostate index $m$. The fermion partition function using the $m$ index is then

\begin{equation}
Z_\text{Fermi}=[t^N]\prod_{j=0}^{d-1}(1+te^{-\gamma j})=[t^N]\prod_{j=0}^{d-1}z_\text{Fermi}(j, t).\label{eq: Fermi partition function}
\end{equation}
The fermion partition function is therefore separable over the macrostates $m$, allowing each state to be treated as a mutually independent grand canonical ensemble, where
$$
\begin{aligned}
z_\text{Fermi}(j, t)&=\sum_{n=0}^1 t^ne^{-n\gamma j}=1+te^{-\gamma j}.
\end{aligned}
$$
The state probability distribution with only 0 or 1 particles in state $j$ is then
$$
P(j,t)=\frac{e^{-\gamma j k_j}}{1+te^{-\gamma j}}
$$
and
$$
P(\mathbf{k}_\text{Fermi})=[t^N]\prod_{j=0}^{d-1}P(j,t),
$$
showing directly that the states $j$ are mutually independent. 
The average number of particles in state $j$ then becomes
$$
\braket{n_j}=\sum_{k_j=0}^1 k_jP(j)=\frac{1}{e^{\gamma j}+1}
$$
which is the Fermi-Dirac distribution \cite{rosser1986fermi}. We have removed the selection variable $t$, as this expression does not impose a constraint on the total particle number $N$. The separability of the probability distribution across the states $j$ allows each state to be treated as mutually independent. However, as in the bosonic case, the particles themselves cannot be considered mutually independent. Consequently, from a statistical perspective, fermions also permit interaction between particles. Finally, in the limit that $\tau\rightarrow\infty$, eq.\eqref{eq: Fermi partition function} becomes
$$
Z_\text{Fermi}=\begin{pmatrix}d\\N\end{pmatrix}.
$$

The entropy of distinguishable, bosonic and fermionic ensembles is presented in Fig.\ref{fig: Comparing distributions}, showing that an ensemble of distinguishable particles can store far more entropy than bosonic and fermionic ensembles. What we hope is clear is that the statistics and probability distributions for energetic systems apply equally to any mutually independent conserved quantity, like spin angular momentum.

\section{Multiplicity of Boson spin states}\label{subsubsec: Multiplicity of indistinugishable particles}

\begin{figure}[t]
    \centering
    \includegraphics[width=0.75\linewidth]{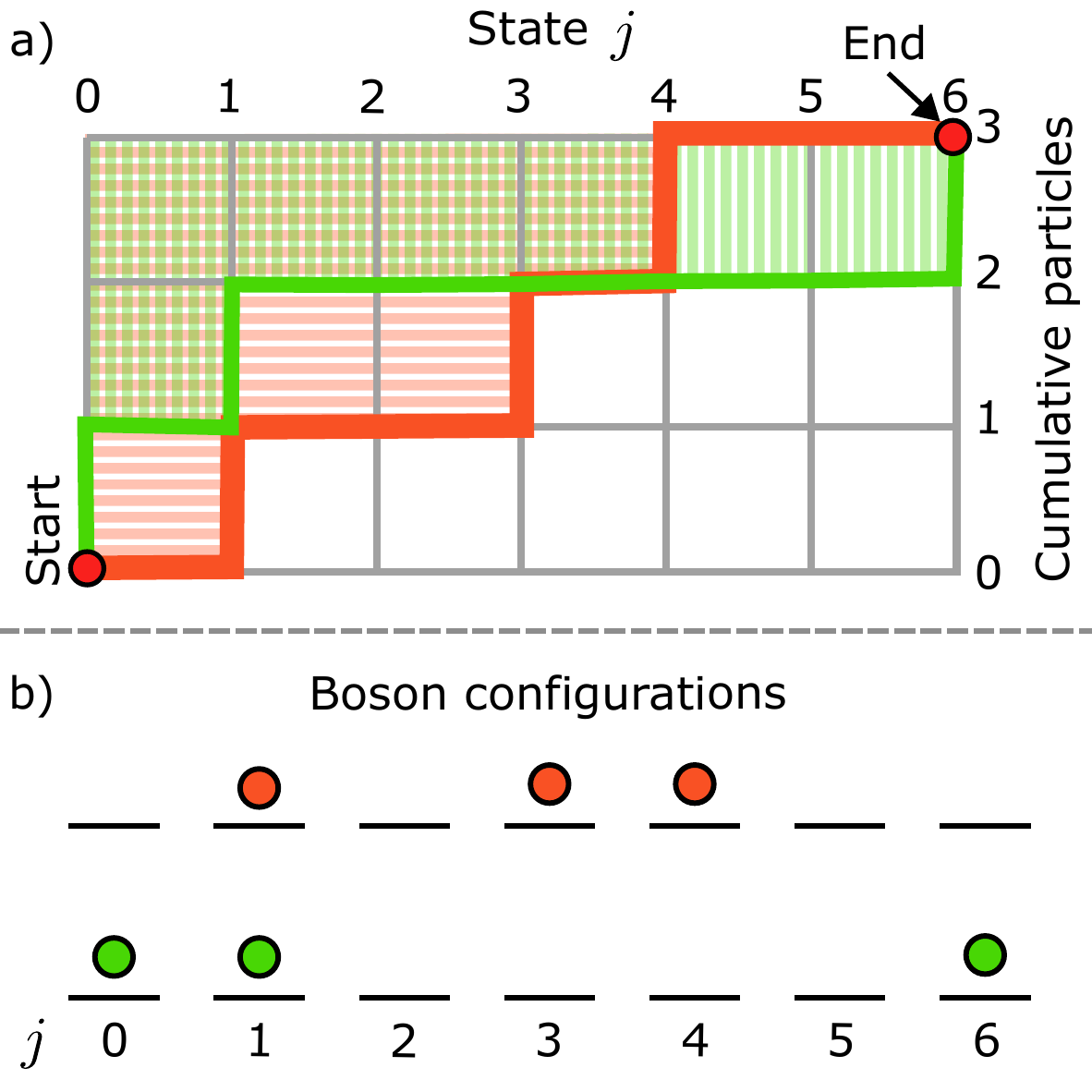}
    \caption{Boson microstates (b) represented using the $N\times d$ grid problem (a), where the grid dimensions correspond to the total particle number $N$ and the number of states per particle $d$. The grid problem counts the number of unique paths starting from the ``Start'' corner and ending at the ``End'' corner, restricted to upward and rightward moves, that result in the same area above the path. In (a) two example paths are shown, with (b) displaying the corresponding boson configurations in the $j$ states. The shaded area represents the macrostate, with areas of 7 and 8 for the green and red paths, respectively, corresponding to the correct macrostates for the particle distributions depicted in (b).}
    \label{fig:Boson grid representation}
\end{figure}
The number of configurations of indistinguishable bosons that yield the same total spin angular momentum can be reformulated as a combinatorial problem: counting the number of unique paths on an $N\times d$ grid, starting from the bottom-left corner and ending at the top-right corner, with movement restricted to upward and rightward steps, as illustrated in Fig.\ref{fig:Boson grid representation}. The grid represents $N$ particles distributed among $d$ states, and the area of the grid above the path taken represents the macrostate $m(\mathbf{k})=\sum_{j=0}^{d-1}j k_j$. 

The question then becomes `how many unique paths result in the same area'? Answering this question is equivalent to determining how many configurations of indistinguishable particles yield the same macrostate. The solution is a generating function of the Gaussian binomial \cite{Azose01052020},
$$
\text{Comb}=[q^m]\begin{pmatrix}N+d-1\\d-1\end{pmatrix}_{q}
$$
where $m=m(\mathbf{k})\in\{0, 1, 2\dots (d-1)N\}$. This can then be interpreted as simply the multiplicity of microstates $g(m)$ for boson when indexed over the macrostates. We can then use this to derive the boson partition function in terms of the macrostates,

$$
Z_\text{Boson}=\begin{pmatrix}N+d-1\\d-1\end{pmatrix}_{q}:\quad q=e^{-\gamma}.
$$

\section{Waste and Entropic responses}\label{sec: Waste capacity}
In thermodynamics, the heat capacity is commonly defined as the amount of heat required to raise the temperature of a system by one unit. Equivalently, it can be interpreted as the amount of waste energy (i.e., heat) that the system can store for a given change in temperature. Heat capacity is defined as
$$
C_V = \left(\frac{\partial U}{\partial T}\right)_{V}
$$
for constant volume $V$ \cite{callen1980thermodynamics}. It can also be expressed in terms of the system’s entropy
$$
C_V=T\left(\frac{\partial \mathbb{S}}{\partial T}\right)_V.
$$
We can define an analogous quantity for the spin degree of freedom. Since spin angular momentum depends only on the spin temperature $\tau$ and has no analog of volume or pressure, we define the ``spin therm capacity'' as

\begin{equation}
C_s=\frac{d \braket{\bar{J}_z}}{d \tau}.\label{eq: Waste capcity slope in terms of angular momentum}
\end{equation}
Using eq.\eqref{eq: Entropy-momentum relation}, we can equivalently write
\begin{equation}
C_s=\tau\frac{d\mathbb{S}}{d\tau}\label{eq: Entropy capacity}
\end{equation}
where we can interpret 
\begin{equation}
\frac{1}{\tau}C_s=\frac{d\mathbb{S}}{d\tau}\label{eq: Waste capacity slope}
\end{equation}
as the amount of entropy the system can store per unit change in spin temperature. It should be clear that any mutually independent conserved quantity possesses an analogous heat capacity. It is therefore natural to define $C_s/\tau$ as the entropic response of the ensemble, a quantity that applies generally to any independent conserved variable, such as spin or energy. $C_s$ should be interpreted as the waste response of the ensemble, specific to the conserved quantity in question and associated with an analogous ``heat'' or the loss of usefulness of that conserved quantity. Re-naming $C_s$ from what is conventionally the `heat' capacity emphasizes that $C_s$ is not a measure of capacity per se, but rather quantifies the absorbed `waste' \textit{in response} to a change in temperature. This is especially evident when considering the expression of the distinguishable waste response from table \ref{table: Thermodynamics of spin systems}. The proportionality to the variance of the conserved quantity is a phenomena found generally in the fluctuation-dissipation theorem \cite{MARCONI_2008}, where $C_s$ is a response function. Renaming the heat capacity to the waste response is therefore justified from this perspective. Eqs.\eqref{eq: Waste capcity slope in terms of angular momentum} and \eqref{eq: Waste capacity slope} are presented as alternate, yet equivalent interpretations of a systems ability to store waste, both from a physical and an entropic perspective. These two equations allow for us to interpret heat engines in \S \ref{sec: The entropy battery} as being machines that either filter conserved quantities by separating the `heat' from the `work', or machines that merely transport entropy and information between the heat baths, with these interpretation also applying to the operation of conventional energy heat engine. The information interpretation of the heat engine is general, and leads naturally to concepts involving information transfer between `waste' baths of different conserved quantities, such as spin and energy. 

\begin{figure}
    \centering
    \includegraphics[width=\linewidth]{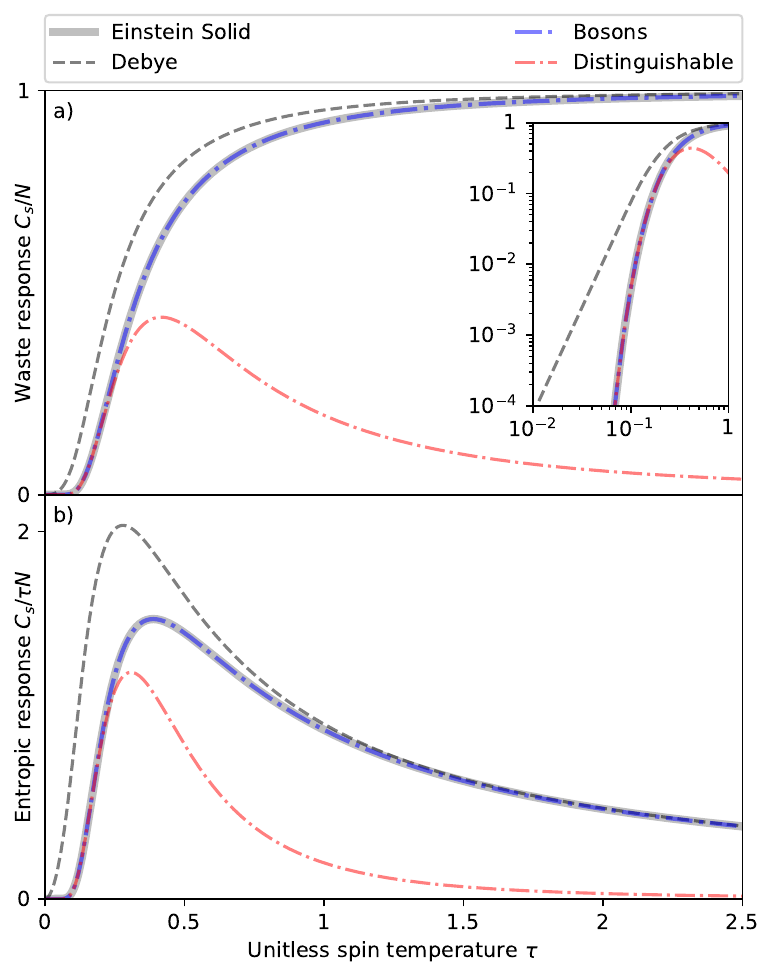}
    \caption{(a) The waste response - colloquially the specific heat capacity - and (b) the entropic response for bosonic and distinguishable ensembles. Also shown are the Einstein solid and Debye models of specific heat capacity for a single degree of freedom \cite{rogers2005einstein}. For all curves $S=1/2$ and the expressions from Table.\ref{table: Thermodynamics of spin systems} were used. The Debye model provides the most accurate description of energetic heat capacity across all temperatures, as it accounts for low-temperature phonon effects commonly observed in metals. The inset in (a) shows the deviation between the Einstein and Debye models in the low-temperature limit and is included for comparison with Fig. 1 of \cite{ashcroft1972lattice}.}
    \label{fig: Einstein solid heat capacity}
\end{figure}

$C_s$ can be expressed generally using eqs.\eqref{eq:AverageSpinInTermsOfZ} and \eqref{eq:Wasteresponse}
\begin{equation}
C_s=2\tau\left(\frac{d}{d\tau}\ln(Z)\right)+\tau^2\left(\frac{d^2}{d\tau^2}\ln(Z)\right)\label{eq: Spin labour capacity},
\end{equation}
showing a dependence only on the partition function. For an ensemble of bosons with infinite particles, where we can treat each state as independent Grand canonical ensembles \eqref{eq: Seperabilty of boson partition function}, then $C_s$ is analytically the specific heat capacity of an Einstein-solid \cite{rogers2005einstein} when $S=\frac{1}{2}$. The waste response is derived for arbitrary spin system composed of bosons and distinguishable particles in Appendix \ref{subsec: Analytic waste response for distinguishable information}-\ref{sec: Waste response for the debye and einstein models of heat capacity}, with the results summarised in Table.\ref{table: Thermodynamics of spin systems}. These results are also presented in Fig.\ref{fig: Einstein solid heat capacity}, along with the unitless, per degree of freedom Debye and Einstein-solid models of specific heat capacity.

The system’s entropy capacity, $\mathbb{S}_\text{max}$, defined as the maximum entropy the system can store, is given by the integral of the entropic response with respect to temperature. From eq.\eqref{eq: Waste capacity slope}

\begin{equation}
\mathbb{S}_{\max}=\int_{0}^{\mathbb{S}_{\max}}d\mathbb{S}=\int_{0}^{\infty}C_s\frac{1}{\tau}d\tau  \label{eq: Entropy max}.
\end{equation}
The waste capacity, defined as the maximum unitless ``heat'' that the ensemble can absorb, can be calculated similarly using eq.\eqref{eq: Spin labour capacity}.
$$
\mathcal{Q}_\text{max}=\int_{-SN}^0 d\braket{\bar{J}_z}=\int_{0}^\infty C_sd\tau
$$
Although trivial in their derivation, these two equations should be understood as representing two distinct ways in which an ensemble can store waste. In one representation, the ensemble stores entropy, corresponding to the lack of information about the system’s distribution across its microstates. In the other, it stores the ``disordered'' physical quantity, heat, or more generally, waste. Evaluating eq.\eqref{eq: Entropy max}, we find that the entropy capacity is
\begin{figure}
    \centering
    \includegraphics[width=\linewidth]{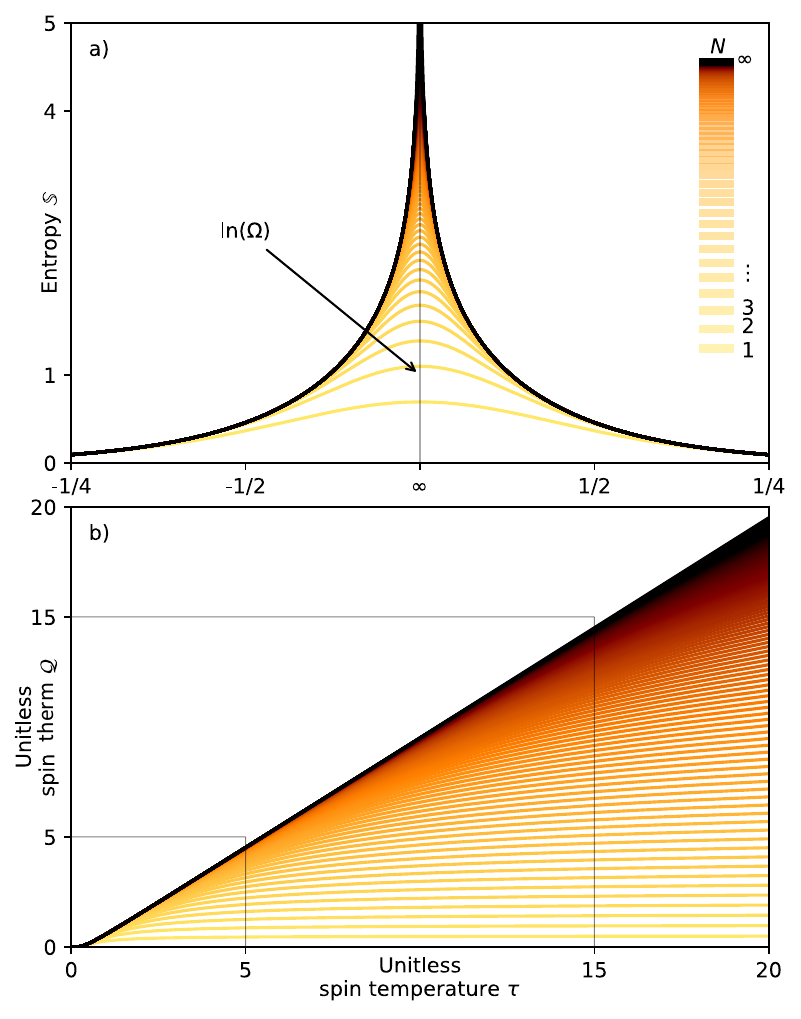}
    \caption{(a) The bosonic entropy and (b) spin therm for varying spin temperature $\tau$ with $S=1/2$. The coloured lines show the numerical entropy and heat for increasing $N$ from $N=1$, calculated using the waste response from eq.\eqref{eq: Spin labour capacity}, $Z_\text{Boson}$ from eq.\eqref{eq: Boson partition function with m index} and integrating eqs.\eqref{eq: Waste capcity slope in terms of angular momentum} and \eqref{eq: Waste capacity slope}. In black in both subplots is the analytic entropy and heat from \eqref{eq: Boson analytic entropy} and \eqref{eq:Bosonic heat}. In (a) we labeled the maximum entropy when $\tau=\infty$ using the grey vertical line, and in (b) we make it clear how in the large $\tau$ limit with large $N$ that $\mathcal{Q}\propto \tau$. (b) also shows the large $\tau$ limit of heat with finite $N$ of $\mathcal{Q}_\text{max}=SN$ from eq.\eqref{eq: Max spin therm}.} 
    \label{fig:analytic boson entropy}
\end{figure}

\begin{equation}
\mathbb{S}_\text{max}=\lim_{\tau\rightarrow\infty}\mathbb{S}=\ln\left(\sum_\Omega 1\right)=\ln(\Omega).\label{eq: Waste capacity}
\end{equation}
$\mathbb{S}_\text{max}$ is therefore determined by the total number of microstates. This is Boltzmann's definition of entropy for energy systems $\mathbb{S}/k_B=\ln(\Omega)$ \cite{perrot1998z}. The entropy for distinguishable particles is extensive with the number of particles, as $|\Omega|=d^N$. However it is non-extensive for indistinguishable particles. The waste capacity or maximum spin therm the spin ensemble can absorb is
\begin{equation}
\mathcal{Q}_\text{max}=SN=\frac{d-1}{2}N\label{eq: Max spin therm}
\end{equation}
which is extensive in the number of particles, regardless of whether they are distinguishable or indistinguishable.

Finally, in the infinite $N$ limit the entropy and spin therm for bosons becomes eqs.\eqref{eq: Boson analytic entropy} and \eqref{eq:Bosonic heat} respectively. The boson spin therm is proportional to $\tau$ in the high temperature limit, as the Taylor expansion of $\coth(1/x)\approx x$ for $x\gg0$, which re-derives the usual heat expression $\mathcal{Q}\propto \tau$ used in, for example, deriving the standard Carnot efficiency limit $\eta=1-\tau_c/\tau_h$. Eqs.\eqref{eq: Boson analytic entropy} and \eqref{eq:Bosonic heat} are presented graphically in Fig.\ref{fig:analytic boson entropy}. 

\section{Analytic waste response}\label{sec: Waste response appendix}
\subsection{Distinguishable particles}\label{subsec: Analytic waste response for distinguishable information}
From eq.\eqref{eq: Distinguishable partition function}, the partition function for an ensemble of $N$ distinguishable particles with $d$ states per particle is

$$
Z_\text{dist}=\left(\sum_{j=0}^{d-1}e^{-j/\tau}\right)^N=Z^N
$$
where we have written the distinguishable partition function in terms of the per particle partition function $Z$. The derivative of $\ln(Z_\text{dist})$ is
$$
\frac{d}{d\tau}\ln(Z_\text{dist})=\frac{N\sum_{j=0}^{d-1}je^{-j/\tau}}{\tau^2\sum_{j=0}^{d-1}e^{-j/\tau}}=\frac{N}{\tau^2}\braket{j(\tau)}
$$
where we have identified the average spin state index $\braket{j(\tau)}$ as being the weighted sum of the states
$$
\braket{j(\tau)}=\sum_{j=0}^{d-1}jP(j, \tau)
$$
for probability $P(j, \tau)=e^{-j/\tau}/Z$, and $Z=\sum_{j=0}^{d-1}e^{-j/\tau}$. Note also that $\braket{j}N=\braket{\bar{J}_z}+SN$. The second derivative is
$$
\frac{d}{d\tau}\frac{N}{\tau^2}\braket{j(\tau)}=-\frac{2N}{\tau^2}\braket{j(\tau)}+\frac{N}{\tau^4}\left(\braket{j(\tau)^2}-\braket{j(\tau)}^2\right).
$$
We can identify the variance in $j$, $\text{Var}(j(\tau))=\braket{j(\tau)^2}-\braket{j(\tau)}^2$, and therefore derive the final form of the waste response
$$
C_s=\frac{N}{\tau^2}\text{Var}(j(\tau))
$$
It can be shown by using
$$
\braket{\bar{J}_z}=\sum_{j=0}^{d-1}m_jP(j)=\sum_{j=0}^{d-1}m_j\frac{e^{-j/\tau}}{Z}
$$
that
$$
\text{Var}(\bar{J}_z(\tau))=\text{Var}(j(\tau)).
$$
This gives us the final form of the per particle waste response for distinguishable information
$$
\frac{C_s}{N}=\frac{1}{\tau^2}\text{Var}(\bar{J}_z(\tau)).
$$

\subsection{Bosons}\label{sec: Instantaneous waste capacity analytic for Bosons}

From appendix \ref{sec: boson case}, the partition function for an ensemble of $N$ boson with $d$ states per particle is
$$
Z_\text{Boson}=\begin{pmatrix}N+d-1\\d-1\end{pmatrix}_p:\quad p=e^{-\gamma}
$$
In the limit that $N\rightarrow\infty$, this becomes
$$
Z_\text{Boson}=\prod_{j=1}^{d-1}\frac{1}{1-e^{-j\gamma}}
$$
from Newton's generalized binomial theorem. We can exploit the fact that the partition function may be multiplied by any constant without altering the normalized probability distribution or the waste response. The choice of constant is motivated by our desire to rewrite $1-e^{-j\gamma}$ into $(e^{j\gamma/2}-e^{-j\gamma/2})/2=\sinh(j\gamma/2)$ for every $j$. Accordingly, we multiply $Z_\text{Boson}$ by 
$$
\prod_{m=1}^{d-1}2^me^{-\gamma m/2}
$$
leaving
$$
Z_\text{Boson}=\prod_{j=1}^{d-1}\frac{1}{\sinh(j/2\tau)}
$$
where we have substituted $\gamma=1/\tau$. Taking the derivative of $Z_\text{Boson}$ with respect to $\tau$, 
$$
\frac{d Z}{d\tau}=\sum_{p=1}^{d-1}\left(\frac{d}{d\tau}\frac{1}{\sinh(p/2\tau)}\right)\frac{1}{\prod_{j=1,j\neq p}^{d-1}\sinh(j/2\tau)},
$$
and dividing by $Z$, we obtain
$$
\begin{aligned}
\frac{d}{d\tau}\ln(Z)=\frac{1}{Z}\frac{dZ}{d\tau}&=\sum_{p=1}^{d-1}\left(\frac{d}{d\tau}\frac{1}{\sinh(p/2\tau)}\right)\sinh(p/2\tau)\\
&=\sum_{j=1}^{d-1}\frac{j\cosh(j/2\tau)}{2\tau^2\sinh^2(j/2\tau)}.
\end{aligned}
$$
We have removed the ``$\text{Boson}$" label for the sake of brevity. The second derivative is
$$
\frac{d^2}{d\tau^2}\ln(Z)=\sum_{j=1}^{d-1}\left[\frac{j^2}{4\tau^4\sinh^2(j/2\tau)}-\frac{j\cosh(j/2\tau)}{\tau^3\sinh(j/2\tau)}\right].
$$
Upon substitution into the waste response $C_s$, we obtain the final form
\begin{equation}
C_s=\sum_{j=1}^{d-1}\frac{j^2}{4\tau^2\sinh^2(j/2\tau)}.\label{eq: Boson heat capacity appendix}
\end{equation}
It should be noted that taking the $N\rightarrow\infty$ limit of the partition function yields the per-particle waste response, as this corresponds to a grand canonical treatment of each spin state.


\subsection{Debye and Einstein models}\label{sec: Waste response for the debye and einstein models of heat capacity}

From \cite{rogers2005einstein} eq.(5.3.2), the Einstein heat capacity is
$$
C_E=3k_BN\left(\frac{h\nu}{k_BT}\right)^2\frac{e^{h\nu/k_BT}}{(e^{h\nu/k_BT}-1)^2}.
$$
This is for 3 degrees of freedom with $N$ particles, energy $h\nu$, and units J/K thanks to the presence of the Boltzmann constant. For 1 degree of freedom, the unitless, per particle heat capacity is
$$
C_E=\left(\frac{h\nu}{k_BT}\right)^2\frac{e^{h\nu/k_BT}}{(e^{h\nu/k_BT}-1)^2}.
$$
The energy levels of the harmonic oscillator are evenly spaced, $h\nu_j=h\nu j$, so that the unitless index is 
$j=h\nu/h\nu_j$, with $h\nu$ corresponding to the $j=1$ level. We also identify the unitless temperature as the Einstein temperature $T_E=h\nu/k_B$, $T/T_E=\tau$, which is equivalent to the unitless spin temperature. The unitless Einstein Solid is therefore
$$
C_E=\frac{1}{\tau^2}\frac{e^{1/\tau}}{(e^{1/\tau}-1)^2}=\frac{1}{\tau^2}\left(\frac{1}{e^{1/2\tau}-e^{-1/2\tau}}\right)^2.
$$

As for the Debye model, the heat capacity is the sum of the individual heat capacities for each $d=2S+1$ states per atom per degree of freedom,

\begin{equation}
C_D=\sum_j^{2S} C_j=\frac{1}{\tau^2}\sum_j^{2S} \left(\frac{j}{e^{j/2\tau}-e^{-j/2\tau}}\right)^2,\label{eq: Boson debye model}
\end{equation}
where we have made the equivalent substitution in the spin degree of freedom, with $2S$ typically being the Debye cut off frequency $\nu_D$. Eq.\eqref{eq: Boson debye model} is equal to the boson model \eqref{eq: Boson heat capacity appendix}. The Debye model approximates the sum using an integral, while including the density of frequency modes \cite{rogers2005einstein}, which is $j^2dj$ in the spin degree of freedom, resulting in the un-normalized Debye model
$$
C_D=\frac{1}{\tau^2}\int_{0}^{2S}\left(\frac{j}{e^{j/2\tau}-e^{-j/2\tau}}\right)^2j^2dj
$$
A normalization factor must be included due to the inclusion of the extra $j^2$ mode density term in the integral. The final unitless Debye model is therefore
$$
C_D=\frac{1}{\tau^2}\frac{3}{(2S)^2}\int_{0}^{2S}\left(\frac{j}{e^{j/2\tau}-e^{-j/2\tau}}\right)^2j^2dj.
$$
This expression equals the per degree of freedom, per particle Debye model from \cite{rogers2005einstein} eq.(6.2.16).

\section{Distinguishability discussion}\label{sec: Distinguishability}
With the goal of this article being to present a realizable entropy storage device, and with entropy capacity translating into extractable energetic work as demonstrated by the entropy battery, then we require an understanding of how one could increase entropy capacity by means other than enlarging the size of the system. As already discussed in \S \ref{sec: Maximising entropy capcity} and appendix \ref{sec: Distinguishable particles}, aside from increasing the number of particles, number of states or the number of independent conserved quantities, the only other apparent means of increasing entropy capacity is to render each particle distinguishable. However, as also established in those sections, a system of distinguishable particles must necessarily be non-interacting. Since particles do in fact interact and heat capacities for distinguishable particles are very different from experimental heat capacities, then particles must be indistinguishable. In this section, we examine whether it is possible to retain the physical entropy capacity associated with distinguishable particles while still allowing interaction. Our conclusion, inspired largely by Saunders’ analysis of indistinguishability \cite{SAUNDERS202037}, is that this is not possible even in principle. The reader is referred to his work for a thorough treatment of the indistinguishability problem.

We begin by clarifying what we mean by distinguishability. From an entropic perspective, physical distinguishability increases the number of accessible microstates $\Omega$ available to the ensemble, leading to an enlarged entropy capacity. Therefore, distinguishability is real when the entropy capacity of the ensemble at a temperature is larger than that of an ensemble of indistinguishable particles at the same temperature. The entropy capacity is determined by the total number of microstates, which in itself is determined by the number of unique distributions of the particles in state space. Under this definition, if two ensemble states are physically identifiable, then these are distinct microstates, with each contributing to the system’s physical entropy and hence to its extractable work.

Consider first a single classical particle with physical properties such as position, momentum, and charge. Let these properties be localised in state space (a minimum entropy state). Since there is only a single particle, then the particle can be tracked continuously through unitary evolution, rendering it distinguishable at all times. Now introduce a second identical particle (same physical properties), however occupying a different location in state space. As long as these particles do not interact and their probability distributions remain separable, they are distinguishable and their entropies add extensively. However, once the particles approach sufficiently closely to interact via mutual forces, can they still be considered distinguishable?

To address this, let us restrict the state space to a single conserved property: linear momentum in one dimension. If a mutual force acts between them, for example, in an elastic collision, momentum is redistributed under the law of conservation of linear momentum. After the interaction, due to this redistribution, the individual particles momenta no longer act as a reliably distinguishable feature. Therefore, in the momentum degree of freedom, the initially distinguishable particles become indistinguishable after collision. Attempts to preserve distinguishability through continuous measurement only introduce further complications. Consider what happens when the two particles are classical. In principle, their properties can be defined with arbitrary precision. However, this implies that the mutual forces responsible for momentum exchange must become arbitrarily small. If any interaction were present, information would be shared between particles, making their individual states interdependent and less distinguishable. In other words, the act of interacting necessarily introduces correlations that blur the precise definition of each particle’s state. Therefore, in the limit of perfect precision, the particles must be effectively non-interacting, just as assumed in the ideal gas model. Moreover, defining distinguishability through measurement is problematic, as measurement must be performed using a probe particle, or via some interaction with the system of interest. The probe must initially be in a known low entropy state, which is at non-equilibrium with the ensemble, and the probe ‘observes’ the microstates of the ensemble through an effective thermalization. The measurement is in effect a temperature measurement for each conserved quantity (see Appendix \ref{sec: Spin temperature}). The probe particle (measurement device) necessarily interacts with the system during the effective thermalization. If the two particles that we intend on distinguishing are sufficiently small particles, the probe will perturb the final momentum distribution, removing the possibility of distinguishing the particles in a subsequent measurement, and ruling out continuous measurement as a means for enabling distinguishability. If the probe is non-perturbative or only weakly interacting, then insufficient information is obtained to guarantee distinguishability of the colliding particles \cite{cheong2012balance}.

The difficulty of persistent distinguishability during the two particle collision continues when additional conserved properties are reintroduced. If each conserved quantity is mutually independent, then interactions only transfer information within their respective state-space planes, with the example of linear momentum in the prior paragraph now applying to each independent conserved quantity. Even when conserved quantities are not mutually independent, interactions still generate mutual information that limits distinguishability in accordance with all conservation laws. Therefore, small classical particles cannot remain distinguishable from an entropic perspective if they interact, even in principle. The fact that particles do interact, and that measured heat capacities are consistent with ensembles of indistinguishable particles \cite{SAUNDERS202037}, reinforces this conclusion.

We are now in a position to propose a general description of particle indistinguishability from the perspective of information thermodynamics. What is it that makes particles fundamentally indistinguishable? Is it the conserved quantities themselves? Is it the particles? We argue that it is neither. Rather, it is information, or coherence in it's most general form that is indistinguishable. This assessment is in agreement with the indistinguishability of information in quantum mechanics, although argued from the perspective of classical mechanics. The generalization of thermodynamics to arbitrary conserved quantities (Appendices \ref{sec: Spin temperature}–\ref{sec: Waste capacity}), combined with the inseparability of probability distributions after interaction, supports this view. It also provides a natural intuition for the coherent transfer of information between mutually independent conserved quantities. It is precisely this redistribution of information that the entropy battery exploits.

\section{Endoreversible efficiency of the entropy battery}
\label{sec: Endoreversible efficiency proof}
In this section we will show that a Carnot like heat engine running between two thermal reservoirs using the models developed in this article satisfy the maximum power efficiency limit for a single endoreversible cycle. With the entropy battery being an extension of these models into the spin angular momentum degree of freedom, then the entropy battery represents the efficiency achievable at maximum power when acting reversibly. 

For this scenario, the reservoirs have temperatures $\tau_{H}, \tau_{C}$ with $\tau_H>\tau_C$. These reservoirs thermalize to a final temperature $\tau_f$ as the unitary heat engines run. The unitless bosonic heat \eqref{eq:Bosonic heat} for $\tau\gg0$ is
$$
Q=\frac{1}{2}\sum_{j=1}^{d-1}(j\coth(j/2\tau)-1)\approx(d-1)(\tau-1/2)
$$
to first order. The heat released by the hot reservoir at temperature $\tau_H$ is therefore
$$
Q_{H}=Q(\tau_f)-Q(\tau_H)=(d-1)\Delta\tau_{f,H},
$$
where $\Delta \tau_{i,j}=\tau_i-\tau_j$. Likewise, the heat absorbed by the cold reservoir at temperature $\tau_C$ is
$$
Q_C=Q(\tau_f)-Q(\tau_C)=(d-1)\Delta\tau_{f,C}
$$
The liberated work is therefore
$$
W=Q_H+Q_C=(d-1)(2\tau_f-\tau_H-\tau_C)
$$
and the efficiency is
$$
\eta=\frac{W}{Q_H}=\frac{2\tau_f-\tau_H-\tau_C}{\tau_f-\tau_C}
$$
For the endoreversible heat engine with no spin reservoir, the final temperature is a standard result $\tau_f=\sqrt{\tau_H\tau_C}$, which is calculated based on the conditions of reversibility $\Delta \mathbb{S}=0$ and eq.\eqref{eq: Entropy-momentum relation}. See \cite{10.1119/1.2397094} for details of the derivation. Upon substitution, and with some algebraic manipulation, we obtain the efficiency
$$
\eta = 1-\sqrt{\frac{\tau_C}{\tau_H}}.
$$
which is the energy efficiency at maximum power \cite{NOVIKOV1958125, 10.1119/1.2397094}. 

Introducing the spin reservoir, the condition of reversibility becomes
\begin{equation}
\Delta \mathbb{S}_H(\tau_f) + \Delta \mathbb{S}_C(\tau_f) + \Delta \mathbb{S}_s(\tau_f) = 0\label{eq: Reversibility condition},
\end{equation}
where we can use the analytic bosonic entropy equation \eqref{eq: Boson analytic entropy} to re-express the changes in entropy
\begin{equation}
\begin{aligned}
\Delta\mathbb{S}_i(\tau_f)&=\sum_{j=1}^{d_i-1}\left(\frac{j}{2\tau_f}\coth(j/2\tau_f)-\frac{j}{2\tau_i}\coth(j/2\tau_i)\right.\\&\left.+\ln\left(\frac{\sinh(j/2\tau_i)}{\sinh(j/2\tau_f)}\right)\right)\label{eq: Delta S analytic}.
\end{aligned}
\end{equation}
This can be approximated to first order for $\tau_i, \tau_f\gg 0$,
\begin{equation}
\Delta\mathbb{S}_i(\tau_f)\approx(d_i-1)\ln\left(\frac{\tau_f}{\tau_i}\right)
\end{equation}
giving an approximate final temperature
\begin{equation}
    \tau_f\approx\left(\tau_{H}^{d_H-1}\tau_C^{d_C-1}\tau_s^{d_s-1}\right)^{1/(d_H+d_C+d_s-3)},\label{eq: Geometric mean of final temperatures}
\end{equation}
which is a weighted geometric mean of the initial temperatures. Setting $d_i$ all equal, then we find $\tau_f\approx\sqrt[3]{\tau_H\tau_C\tau_s}$, which is consistent with the final temperature of a three body system, each being the same size, yet unequal temperatures \cite{10.1119/1.2397094}. The implication of this result is that the energy efficiency obtained in \S \ref{sec: The entropy battery} is for endoreversible engines running at maximum power. However this result does not determine the actual power of the engine. As such, we make no claim as to the actual power of the entropy battery, the analysis of which would require finite-time thermodynamics \cite{e22101076}. We therefore consider the entropy battery as operating at maximum power assuming it runs slowly and isolated enough to fulfill the condition of reversibility \eqref{eq: Reversibility condition}.

\end{document}